\documentclass[a4paper,aps,onecolumn,nofootinbib,preprintnumbers,superscriptaddress,amsmath,amssymb,amsfonts]{revtex4}
\usepackage{graphicx}
\usepackage{bm,natbib}
\usepackage{amsmath}
\usepackage[english]{babel}
\usepackage[T1]{fontenc}
\usepackage{amssymb}
\usepackage{amsfonts}
\usepackage{epsfig}
\usepackage{colordvi}
\usepackage{psfrag}
\usepackage{color}
\usepackage{ mathrsfs }
\newcommand{\Lagr}{\mathcal{L}}

\newcommand{\G}{\mathcal{G}}

\newcommand{\R}{\mathcal{R}}
\newcommand{\La}{\mathscr{L}}
\usepackage{dcolumn}
\usepackage{multirow}
\usepackage{hyperref}
\usepackage{epstopdf}
\usepackage{bm}
\usepackage{rotating}
\usepackage{lscape}
\usepackage{times}
\usepackage{comment}

\hypersetup{
  colorlinks=true,        
  linkcolor=blue,         
  citecolor=magenta,      
}

\begin{document}
\title{Exact Solutions in Higher-Dimensional Lovelock and $AdS_5$ Chern--Simons Gravity}

\author{Francesco Bajardi}
\email{francesco.bajardi@unina.it}
\affiliation{Department of Physics ``E. Pancini'', University of Naples ``Federico II'', Naples, Italy.}
\affiliation{INFN Sez. di Napoli, Compl. Univ. di Monte S. Angelo, Edificio G, Via Cinthia, I-80126, Naples, Italy.}

\author{Daniele Vernieri}
\email{daniele.vernieri@unina.it}
\affiliation{Department of Physics ``E. Pancini'', University of Naples ``Federico II'', Naples, Italy.}
\affiliation{INFN Sez. di Napoli, Compl. Univ. di Monte S. Angelo, Edificio G, Via	Cinthia, I-80126, Naples, Italy.}

\author{Salvatore Capozziello}
\email{capozziello@na.infn.it}
\affiliation{Department of Physics ``E. Pancini'', University of Naples ``Federico II'', Naples, Italy.}
\affiliation{INFN Sez. di Napoli, Compl. Univ. di Monte S. Angelo, Edificio G, Via Cinthia, I-80126, Naples, Italy.}
\affiliation{Scuola Superiore Meridionale, Largo San Marcellino 10, I-80138, Naples, Italy.}
\affiliation{Laboratory for Theoretical Cosmology, Tomsk State University of Control Systems and Radioelectronics (TUSUR), 634050 Tomsk, Russia.}

\date{\today}

\begin{abstract}
Lovelock gravity in $D$-dimensional space-times is considered adopting Cartan's structure equations.  In this context, we find out exact solutions in cosmological and spherically symmetric backgrounds. In the latter case, we also derive horizons and the corresponding Bekenstein--Hawking entropies. Moreover, we focus on the topological Chern--Simons theory, providing exact solutions in 5 dimensions. Specifically, it is possible to show that Anti-de Sitter invariant Chern--Simons gravity can be framed within Lovelock--Zumino gravity in 5 dimensions, for particular choices of Lovelock parameters. 
\end{abstract}

\maketitle

\section{Introduction}

In the last decades, experiments and observations pointed out several phenomena which cannot be framed in General Relativity (GR) at ultraviolet and infrared scales. In particular, most of these issues show incompatibilities of GR at cosmological and astrophysical scales. From one side, GR perfectly fits  observations at Solar System scales, but on the other hand it is not able to give an exhaustive and self-consistent picture  of phenomena like the  acceleration of the late-time universe, so far addressed to the vague concept of Dark Energy. Similarly, inconsistencies in the galaxy rotation curves are relied to Dark Matter which has been, up to now, never detected under the standard of some new  fundamental particle beyond the known families. These are just two representative examples, but the problems occurring once fitting the theory with the large-scale structure observations are several~\cite{Joyce:2014kja, Koyama:2015vza, Capozziello:2010zz}. In light of these issues, alternative theories to GR have been proposed with the purpose of describing the gravitational interaction including extra-terms or modifications into the Einstein--Hilbert action. For instance, a straightforward extension is provided by the $f(R)$  gravity~\cite{Capozziello2002, Capozziello:2009nq, Nojiri:2017qvx, Capozziello:2011et, Capozziello:2015hra, Ribeiro:2021gds}, where the starting action is a general function of the scalar curvature. More generally, functions of higher-order curvature invariants can play the role of the Ricci scalar into the action, providing exhaustive explanations also in the small-scale regime, that is at ultraviolet scales~\cite{Capozziello:2014ioa, Terrucha:2019jpm, Barros:2019pvc, Bajardi:2020mdp, Blazquez-Salcedo:2016enn, Blazquez-Salcedo:2017txk}. Several open issues, involving the early and late-time acceleration of the Universe, can be solved by coupling the geometry to a scalar field $\phi$, as discussed \emph{e.g.} in Refs.~\cite{Amendola:1999qq, Uzan:1999ch, Halliwell:1986ja, Bajardi:2020xfj}. Modified actions give rise to  effective energy-momentum tensors of the gravitational field,  mimicking  the phenomenology observed for Dark Matter and Dark Energy~\cite{Capozziello:2007ec, Clifton:2011jh, Bamba:2012cp, Nojiri:2017ncd, Capozziello:2019klx, Mantica}.

At very small scales, GR cannot be considered by using the same standard as other field theories. Indeed this leads to several shortcomings towards the construction of a self-consistent quantum theory of gravity. First of all, in order to merge the formalism of GR with other interactions of Standard Model, the former must be recast as a Yang--Mills theory. Moreover, even the 2-loop expansion of the gravitational action shows that incurable divergences occur in GR, which cannot be renormalized through the standard regularization procedure. Despite many attempts to merge GR with Quantum Mechanics have been made so far, a complete theory of Quantum Gravity is still missing. One of the first quantization procedure  has been  the so-called ``Arnowitt--Deser--Misner'' (ADM) formalism which, dealing with an infinite dimensional superspace, leads to a Schroedinger-like equation called Wheeler--DeWitt equation~\cite{Bajardi:2020fxh, Hartle:1983ai, Hawking:1983hj}. The ADM formalism does not account for a  full quantum theory of gravity and contains several shortcomings that cannot be overcome.

A forthcoming theory, aimed at solving high-energy issues, should deal with  gravitational interaction as a gauge theory, since gauge theories are, up to now,  the only candidates capable of selecting renormalizable quantum field theories. An example of gauge theory of gravity, dealing with a flat tangent space-time, is the so called \emph{Teleparallel Gravity}, which is invariant under the local translation group and whose action differs from Einstein--Hilbert one for a total divergence. It describes gravity through a torsional space-time, including the antisymmetric contribution of Christoffel connections. The gravitational field is given by the {\it vielbiens} (tetrads in 4 dimensions) and the affinities related  to them are given by the Weitzenb\"ock connection. For a detailed discussion on Teleparallel Gravity and its applications see \emph{e.g.} Refs.~\cite{Cai:2015emx, Ferraro:2006jd, Arcos:2005ec, Bajardi:2021tul}. 

An other generalization of GR was introduced by Lovelock in 1971~\cite{Lovelock:1971yv}. In such a theory, the Lagrangian (the Lovelock--Zumino Lagrangian or, simply, the Lovelock Lagrangian) is the most general torsionless Lagrangian leading to second-order field equations. As better pointed out in the next Sections, the 3-dimensional Lovelock Lagrangian is only made by a sum of the Ricci scalar and the cosmological constant. In 4 dimensions also the Gauss--Bonnet terms arises, though it does not provide any contributions to the equations of motion. As a matter of fact, in 4 dimensions, the Gauss--Bonnet term turns into a topological surface term and starts being non-trivial in 5 dimensions. Any Lovelock Lagrangian, regardless of the dimension considered, is always invariant at least under the local Lorentz group by construction. However, some particular combinations of the coupling constants make the theory invariant with respect to other gauge groups. An exception is given by the Lovelock 3-dimensional Lagrangian, which is invariant under the local Poincare group for any coupling parameter. Moreover, it turns out that there is a particular subclass of Lovelock Lagrangians whose coupling constants are combined such that the external derivative of the Lagrangian provides a topological invariant. Such particular Lagrangians are called ``Chern--Simons Lagrangians'' and provide non-trivial contributions to the dynamics only in odd dimensions. 

In this paper, after dealing with the general $D$-dimensional Lovelock action, we focus on the Chern--Simons 5-dimensional theory of gravity (a particular case of Lovelock gravity), invariant under the local Anti-de Sitter (AdS) group, deriving exact solutions both in cosmology and in spherical symmetry. As discussed in Refs.~\cite{Zanelli:2005sa, Achucarro:1987vz, Chapline:1982ww, Benna:2008zy, Chamseddine:1990gk}, the Chern--Simons theory is mostly used in String Theory and in Supergravity as a starting point for a full  Quantum Gravity theory. In the cosmological framework, Chern--Simons gravity enlarges the predictions provided by GR in higher dimensions~\cite{Aviles:2016hnm, Gomez:2011zzd}.

In the following, we will use the formalism of Vielbein fields, a mathematical tool which links the Minkowski flat space-time and the curved space-time by means of the relation $g_{\mu \nu} = e^a_{\mu} e^b_{\nu} \eta_{a b}$ (we use ``anholonomic'' Latin indices to label tangent flat space-time and ``holonomic'' Greek indices to label curved space-time).

The search for exact solutions, both in cosmology and in spherical symmetry, is pursued by means of the Lagrangian approach. In this regard, we want to point out that, generally, Lagrangian approach and variational approach do not provide the same solutions. However, in this case, the set of Euler--Lagrange equations turns out to be equivalent to the field equations obtained by means of the variational approach.  In cosmology, the first Friedmann equation is recovered by means of the energy condition, while the second field equation is the Euler--Lagrange equation with respect to the scale factor. In spherical symmetry, the variational approach provides three equations, which are not linearly independent. Therefore, the dynamics is provided by two field equations only. Such two field equations are exactly equivalent to the Euler--Lagrange equations with respect to $P(r)$ and $Q(r)$, \emph{i.e.} the two functions in the spherically symmetric line element. Therefore, in the present case, the Lagrangian approach and the variational approach turn out to be exactly equivalent.

The paper is organized as follows: in Sec.~\ref{Lovelock and Chern--Simons Gravity} we outline the main features of Lovelock and Chern--Simons gravity. Secs.~\ref{LOVCOS} and~\ref{CSSS} are devoted to the search for exact solutions in  cosmological and  spherically symmetric backgrounds, respectively. Moreover, we find out exact solutions in $5$ dimensions and generalize the results to $d+1$ dimensions. Finally, in Sec.~\ref{Concl} we conclude and discuss the results. 

Hereafter  we  use natural units: $\hbar = c = k_B = G_N = e^+ = 1$. Moreover, $D$ labels the space-time dimensions, and $d$ refers to the spatial dimensions only.

\section{Lovelock and Chern--Simons Gravity}
\label{Lovelock and Chern--Simons Gravity}

Two of the most discussed assumptions of GR are  the request of symmetric connection to label the geodesic structure and the fact that dynamics is endowed with  second-order field equations. When the former is relaxed and Christoffel connection is not assumed to be torsionless \emph{a priori}, the so called ``Einstein--Cartan formalism'' can be developed. The latter relies on the choice of the  gravitational field action, linear in the Ricci scalar  which, once extended, can lead to modified theories of gravity with higher-order field equations. The class of theories considered here deals with second-order field equations without assuming a symmetric connection. 

At the fundamental level, the invariance of the theory with respect to general coordinate transformations yields several problems in constructing a theory of Quantum Gravity; diffeomorphism invariance in fact involves local coordinates, while gauge invariance deals with field variation. The first step towards a quantum formulation, is to treat gravity under the standard of the  other interactions, where the dynamical fields evolve in a flat space-time. With the aim to connect the curved space-time with the local tangent space, the Einstein--Cart\'an formalism has been formulated. It adopts a mathematical tool capable of relating a general metric tensor to the Minkowski tensor by means of the relation:
\begin{equation}
g_{\mu \nu} = e^a_{\mu} e^b_{\nu} \eta_{a b} \nonumber \;.
\end{equation}   
In order to develop a $D$-dimensional gauge theory of gravity in the tangent space-time, which is invariant under the local Lorentz group, the action must be constructed by means of curvature, torsion, Levi--Civita symbol and Minkowski metric, which are all invariants with respect to $SO(1,D-1)$. In this way, the 4-dimensional Einstein--Hilbert action, invariant under the local Lorentz group, takes the form \cite{Regge:1986nz}:
\begin{equation}
S_{E-H} = \epsilon_{abcd}\int \frac{1}{16 \pi} R^{ab} \wedge e^c \wedge e^d + \Lambda e^a \wedge e^b \wedge e^c \wedge e^d\,,
\label{azione he lorentz invariante}
\end{equation}
where $\Lambda$ is the cosmological constant and $e^a$ is the set of basis defined through the tetrad fields as
\begin{equation}
e^a = e^a_{\mu} dx^\mu\,.
\end{equation}
Einstein's field equations are thus recovered by varying the action with respect to the basis $e^a$. However,in higher dimensions, the action in Eq.~\eqref{azione he lorentz invariante} can be further extended to a more general Lorentz invariant action. Specifically, the most general $D$-dimensional action which does not manifestly contain torsion, invariant under local Lorentz transformations, and leading to second-order field equations is~\cite{Lovelock:1971yv, Cvetkovic:2016ios, Mardones:1990qc, Exirifard:2007da}
\begin{equation}
S = \kappa \int_\mathcal{M} \sum_{i=0}^{\frac{D}{2}} \alpha_i \La^{(D,i)}, 
\label{d-dim lovelock}
\end{equation}
with $\kappa$ being a dimensionless constant and where $\La^{(D,i)}$ is defined as:
\begin{equation}
\La^{(D,i)} = \epsilon_{a_1,a_2...a_D} R^{a_1 a_2} \wedge R^{a_3 a_4} \wedge ... \wedge R^{a_{2i-1} a_{2i}} \wedge e^{a_{2i+1}} \wedge e^{a_{2i+2}} \wedge ... \wedge e^{a_D} \;.
\label{Lagrangiana di Lovelock}
\end{equation}
The Lagrangian in Eq.~\eqref{Lagrangiana di Lovelock} is the so-called \emph{Lovelock Lagrangian}. Notice, however, that albeit torsion does not appear in the Lovelock Lagrangian, assuming torsion to vanish is generally unnecessary. Torsion may appear after performing the exterior derivative or after varying the Lagrangian with respect to the two-form connection $\omega^{ab}$ (see Appendix~\ref{APPA} for details). However, in this paper, we make computations under the assumption of zero torsion. Consequently, the two-form curvature $R^{ab}$ can be written in terms of the one-form basis $e^a$ only, as it can be noticed from Eqs.~\eqref{structural eq1} and~\eqref{structural eq2}. Setting \emph{e.g.} $D=4$, the Lovelock action reduces to
\begin{equation}
S_L = \kappa \int_\mathcal{M} \sum_{i=0}^{2} \alpha_i \La^{(4,i)} = \kappa \int \epsilon_{abcd} \left( \alpha_0 e^a \wedge e^b \wedge e^c \wedge e^d + \alpha_1 R^{ab} \wedge e^c \wedge e^d + \alpha_2 R^{ab} \wedge R^{cd} \right),
\label{Lovelock action}
\end{equation}
containing the scalar curvature and the Gauss--Bonnet term $\epsilon_{abcd} R^{ab} \wedge R^{cd}$ only. As mentioned above, in 4 dimensions, the Gauss--Bonnet scalar does not provide any contribution to the dynamics, thus the variation of the action in Eq.~\eqref{Lovelock action} with respect to the basis $e^a$ yields exactly the Einstein equations with cosmological constant. Basic foundations and applications of Lovelock gravity have been extensively studied in the literature, \emph{e.g.} in Ref.~\cite{Deruelle:2003ck}, where the Authors consider the \emph{quasi}-linearity of Lovelock field equations by means of Kaluza--Klein and brane cosmological models. In Ref.~\cite{Boulware:1985wk}, the Authors consider Lovelock Lagrangian as coming from the expansion of supersymmetric string theory, showing that this model has both flat and AdS space as solutions. They also study the stability of cosmological and spherically symmetric solutions, discussing the effects of higher-order corrections to the Einstein--Hilbert action. In Ref.~\cite{Castillo-Felisola:2016kpe}, cosmological solutions of 5-dimensional Lovelock gravity in the dimensionally reduced theory are provided, and the presence of torsion yields a non-vanishing energy-momentum tensor in four dimensions. In Ref. \cite{Teitelboim:1987zz}, the authors construct the Hamiltonian formulation of Lovelock gravity in higher dimensions.

In 3 dimensions, the Lagrangian in Eq.~\eqref{Lagrangiana di Lovelock} turns out to be:
\begin{equation}
\La_3^L=\epsilon_{abc} \left( \alpha_0 e^a \wedge e^b \wedge e^c + \alpha_1 R^{ab} \wedge e^c  \right)\,.
\label{Lovelock 3D}
\end{equation}
Notice that when $\alpha_0 = 0$, Eq.~\eqref{Lovelock 3D} is invariant with respect to the local Poincar\'e group. From the above Lagrangian, it is possible to obtain an important property shared by all Lagrangians which are invariant with respect to the local internal transformations in the flat space-time. It is worth noticing, in fact, that the exterior derivative of Eq.~\eqref{Lovelock 3D} is a topological surface term, specifically the $3+1-D$ Euler density. In general, all Lagrangians whose exterior derivative provides a topological term, are \emph{quasi}-gauge-invariant, namely they only change by a total derivative after performing a gauge transformation. 

Reversing the argument, this means that $D+1$-dimensional topologically invariant forms, can be used to define $D-$forms which in turn provides suitable Lagrangians. These forms are called \emph{Chern--Simons forms} and can be defined in odd dimensions only. However, the lack of non-trivial topological surface terms in odd dimensions, yields the impossibility of constructing non-trivial gauge Lagrangians in even dimensions. For example, the Chern--Simons 3-form coming from the 4-dimensional Pontryagin density $R^{ab} R_{ab}$ is:
\begin{equation}
CS_{3}^{P} = \omega^a_b d \omega^b_a + \frac{2}{3} \omega^a_b \omega^b_c \omega^c_a\,,
\end{equation}
being $\omega^a_b$ the 1-form Lorentz connection defined in Appendix~\ref{APPA}. Another example is given by the Euler density, whose $2D$-dimensional representation is
\begin{equation}
E_{2D} = \epsilon_{a_1 a_2 ... a_{2D-1} a_{2D}} R^{a_1 a_2} \wedge ... \wedge R^{a_{2D-1} a_{2D}}\,,
\end{equation}
and in four dimensions is equivalent to the Gauss--Bonnet term. Moreover, the $2D$-dimensional Euler density turns out to be the exterior derivative of the Chern--Simons Lagrangian invariant with respect to the AdS group, namely:
\begin{equation}
\La^{AdS}_{2D-1} = \sum_{i=0}^{D-1} \tilde{\alpha}_i \La^{(2D-1,i)}\,, \qquad \;\;\; \tilde{\alpha}_i = \frac{(\pm 1)^{i+1} l^{2i-D}}{D-2i} \binom{D-1}{i}\,,
\label{AdS2D-1}
\end{equation}
Lovelock gravity with AdS asymptotics is discussed in detail \emph{e.g.} in Refs.~\cite{Miskovic:2007mg, Miskovic:2010ui, Kofinas:2008ub}. 

In this paper, we mainly focus on the 5-dimensional Lovelock Lagrangian, with particular interest in those coupling constants yielding the invariance under the local AdS group. In this perspective, the 5-dimensional Chern--Simons Lagrangian is
\begin{equation}
\La^{AdS}_{5} = \frac{1}{l} \epsilon_{abcde} \left(R^{ab} \land  R^{cd} \land e^e + \frac{2}{3 l^2} R^{ab} \land e^c \land e^d \land e^e + \frac{1}{5 l^4} e^a \land e^b \land e^c \land e^d \land e^e \right)\,,
\label{CS5}
\end{equation} 
whose exterior derivative provides the following 6-D Euler density~\cite{Oliva:2010zd}:
\begin{eqnarray}
E_6 &=& 2 R^{abcd}  R_{cdef}  R^{ef}_{ab} + 8 R^{ab}_{cd}  R^{ce}_{bf}   R^{df}_{ae} + 24 R^{abcd}  R_{cdbe}  R^e_a + 3 R  R^{abcd}  R_{abcd}  \nonumber
\\
&&+ 24 R^{abcd}  R_{ac}  R_{bd} + 16 R^{ab} R_{bc}  R^c_a - 12 R  R^{ab}  R_{ab} + R^3 \,.
\end{eqnarray}
In Eq. \eqref{CS5}, $l$ is a constant with dimension of length. Notice that the Lagrangian in Eq.~\eqref{CS5} is contained in the general 5-dimensional Lovelock Lagrangian
\begin{equation}
\La^L_{5} =  \epsilon_{abcde} \left( \alpha_0 \; e^a \wedge e^b \wedge e^c \wedge e^d \wedge e^e + \alpha_1 R^{ab} \wedge e^c \wedge e^d \wedge e^e + \alpha_2 R^{ab} \land R^{cd} \wedge e^e \right)\,,
\label{Lovelock5D}
\end{equation}
with $\alpha_0$, $\alpha_1$, $\alpha_2$ defined according to the relation~\eqref{AdS2D-1}. Notice that the study of the 5-dimensional AdS-invariant Chern--Simons Lagrangians is relevant for several reasons. On the one hand, unlike the Poincar\'e group, AdS group is semi-simple, which means that it can be recast as a direct sum of simple Lie algebras. Moreover, according to Cartan's criteria, semi-simple groups are closely related to non-degenerate Killing forms, which in turn can be used to define kinetic terms for the gauge fields and make AdS Lagrangians physically worth for gravity at small scales. Moreover, 5-dimensional (or even higher-dimensional) theories of gravity are largely considered with the purpose to develop a theory that unifies the 4 fundamental interactions. This is the case \emph{e.g.} of String Theory~\cite{Green:1987sp, Green:1987mn, Polchinski:1998rq, Polchinski:1998rr, Becker:2007zj}, $M$-Theory~\cite{Witten:1995em, Dine:2000bf}, Kaluza--Klein theory~\cite{Klein:1926tv, Kaluza:1921tu, Han:1998sg, Mueller-Hoissen:1985www}, \emph{etc}. 

The existence of higher dimensions was firstly conjectured by Einstein himself, who suggested that electromagnetism could be the result of a gravitational field polarized in the fifth dimension. This prescription, however, requires a modification of GR, because the Minkowski space-time and the Maxwell equations are supposed to be embedded in a 5-dimensional space endowed with a Riemann curvature tensor. 

Though several attempts to detect the presence of extra dimensions have been pursued, so far there is no evidence for their occurrence. However, indications of their existences can be provided indirectly. This is the case \emph{e.g.} of the holographic principle, according to which the extra dimensions are manifestations of curvature effects in a space-time with one fewer dimension. 

Though in this work we will mainly consider the AdS-invariant Chern--Simons action as a particular case of the 5-dimensional Lovelock action, it is worth remarking that, from Eq.~\eqref{Lagrangiana di Lovelock}, it is possible to consider  other kinds of gauge Lagrangians, belonging to different gauge groups. Most of them are mainly studied  in Supersymmetry and String Theory. In the next Section, we start from Eq.~\eqref{CS5} and Eq.~\eqref{Lovelock5D} and find out exact solutions of the field equations in a Friedmann--Lema\^itre--Robertson--Walker (FLRW) space-time. Subsequently, in Sec.~\ref{CSSS}, we consider a spherically symmetric background, outlining the corresponding field equation solutions.

\section{Lovelock and $AdS_5$ Chern--Simons Gravity in Cosmology}
\label{LOVCOS}

In more than 4 dimensions, the FLRW metric can be extended by including diagonal time-dependent terms. These terms account for new scale factors which, in general, are different than that related to the standard spatial dimensions. We start by assuming that the 5-dimensional space-time is not isotropic, extending the FLRW metric to:
\begin{equation}
ds^2 = (dx^0)^2 - a^2(t)\left[(dx^1)^2+(dx^2)^2+(dx^3)^2\right] - b^2(t) (dx^4)^2\,,
\label{LEFLRW}
\end{equation}
where $x^4$ is the fifth dimension, labeled by the scale factor $b(t)$. In order to find a solution for the field equations in the extended FLRW interval, we use the first Cartan structure equation to firstly get the 2-form curvature. 
Considering the line element in Eq.~\eqref{LEFLRW}, the set of basis can be chosen as:
\begin{equation}
e^0 = dx^0\,, \;\;\; e^1 =  -a(t) dx^1\,, \;\;\; e^2 = -a(t) dx^2\,, \;\;\;  e^3 =-a(t) dx^3\,, \;\;\; e^4 = -b(t) dx^4\,,
\end{equation}
with a set of tetrad fields of the form
\begin{equation}
e^a_\mu = \text{diag} \left(1, -a(t), -a(t), - a(t), -b(t) \right).
\end{equation}
From Eq.~\eqref{external derivative}, it follows that the exterior derivatives of the set of independent vectors are:
\begin{eqnarray}
&&de^0 = 0, \;\;\;\; de^1 =  \partial_0 (e^1) \land dx^0 = - \dot{a}(t) dx^1 \land dx^0 = \frac{\dot{a}}{a} e^1  \land e^0,  \;\;\;\; de^2  =  \partial_0 (e^2) \land dx^0 = -\dot{a}(t) dx^2  \land dx^0 = \frac{\dot{a}}{a} e^2  \land e^0\,, \nonumber
\\
&& de^3 =  \partial_0 (e^3) \land dx^0 = -\dot{a}(t) dx^3 \land dx^0 = \frac{\dot{a}}{a} e^3  \land e^0\,,  \;\;\;\;\;\; de^4  =  \partial_0 (e^4) \land dx^0 = -\dot{b}(t) dx^4 \land dx^0 = \frac{\dot{b}}{b} e^4 \land e^0\,.
\label{tetrads}
\end{eqnarray}
Assuming the absence of torsion and of extra bosonic fields, the second Cartan structure Eq.~\eqref{structural eq2} permits to find the Lorentz connection
\begin{equation}
\omega^1_0 = \frac{\dot{a}}{a} e^1\,,\;\;\;\; \omega^2_0 = \frac{\dot{a}}{a} e^2\,,\;\;\;\; \omega^3_0 = \frac{\dot{a}}{a} e^3\,,\;\;\;\; \omega^4_0 = \frac{\dot{b}}{b} e^4\,,\;\;\;\;\; \omega^0_i = \omega^i_0\,, 
\end{equation}
and its exterior derivatives
\begin{eqnarray}
d\omega^1_0 &=& e^1 d \left( \frac{\dot{a}}{a} \right) + \frac{\dot{a}}{a} de^1= \left[ \left(\frac{\ddot{a}}{a} - \frac{\dot{a}^2}{a^2} \right) + \frac{\dot{a}^2}{a^2} \right] e^1 \land e^0 = \frac{\ddot{a}}{a} e^1 \land e^0\,, \nonumber
\\
&& d\omega^2_0 = \frac{\ddot{a}}{a} e^2 \land e^0\,, \;\;\;\;\; d\omega^3_0 = \frac{\ddot{a}}{a} e^3 \land e^0\,, \;\;\;\;\; d\omega^4_0 = \frac{\ddot{b}}{b} e^4 \land e^0\,.
\end{eqnarray}
Using Eq.~\eqref{structural eq2}, we finally get the curvature 2-form $R^{ab}$:
\begin{equation}
R^{ab} 
= \begin{pmatrix} \displaystyle
 0 & \frac{\ddot{a}}{a} e^0 \land e^1 & \frac{\ddot{a}}{a} e^0 \land e^2 &  \frac{\ddot{a}}{a} e^0 \land e^3 & \frac{\ddot{b}}{b} e^0 \land e^4 \\ \\ 
 - \frac{\ddot{a}}{a} e^0 \land e^1 & 0 & \frac{\dot{a}^2}{a^2} e^1 \land e^2 & \frac{\dot{a}^2}{a^2} e^1 \land e^3 & \frac{\dot{a} \dot{b}}{ab} e^1 \land e^4 \\ \\
 - \frac{\ddot{a}}{a} e^0 \land e^2 & - \frac{\dot{a}^2}{a^2} e^1 \land e^2 & 0 & \frac{\dot{a}^2}{a^2} e^2 \land e^3 & \frac{\dot{a} \dot{b}}{a b} e^2 \land e^4 \\ \\
 - \frac{\ddot{a}}{a} e^0 \land e^3 & -\frac{\dot{a}^2}{a^2} e^1 \land e^3 & - \frac{\dot{a}^2}{a^2} e^2 \land e^3 & 0 & \frac{\dot{a} \dot{b}}{ab} e^3 \land e^4  \\ \\
 - \frac{\ddot{b}}{b} e^0 \land e^4 & -\frac{\dot{a} \dot{b}}{ab} e^1 \land e^4  & - \frac{\dot{a} \dot{b}}{a b} e^2 \land e^4 & -\frac{\dot{a} \dot{b}}{ab} e^3 \land e^4 & 0   \;
\end{pmatrix}.
\label{curvature}
\end{equation}

Let us now focus on the 5-dimensional limit of the Lovelock action~\eqref{d-dim lovelock}, namely
\begin{equation}
S = \kappa \int \epsilon_{abcde} \left(\alpha_0 e^a \land e^b \land e^c \land e^d \land e^e + \alpha_1  R^{ab} \land e^c \land e^d \land e^e + \alpha_2 R^{ab} \land  R^{cd} \land e^e \right)\,.
\label{startact1}
\end{equation}
By means of the 2-form curvature components in Eq.~\eqref{curvature}, we obtain:
\begin{equation}
 \epsilon_{abcde} R^{ab} \land  R^{cd} \land e^e =\left[ 24\frac{\dot{a}^2 \ddot{a}}{a^3} + 48 \frac{\ddot{a} \dot{a}}{a^2} \frac{\dot{b}}{b} + 24 \frac{\dot{a}^2}{a^2} \frac{\ddot{b}}{b} + 24 \frac{\dot{a}^3}{a^3} \frac{\dot{b}}{b} \right] e^0 \land e^1 \land e^2 \land e^3 \land e^4 \equiv [ \mathcal{G}^{(5)} ] \, e^0 \land e^1 \land e^2 \land e^3 \land e^4\,,
\label{5D GB}
\end{equation}
and
\begin{equation}
\epsilon_{abcde} R^{ab} \land e^c \land e^d \land e^e  = -\left[6 \left(\frac{\ddot{a}}{a} + \frac{\dot{a}^2}{a^2} \right) + 2 \frac{\ddot{b}}{b} + 6 \frac{\dot{a}}{a} \frac{\dot{b}}{b}\right] e^0 \land e^1 \land e^2 \land e^3 \land e^4 \equiv [ \mathcal{R}^{(5)} ] \, e^0 \land e^1 \land e^2 \land e^3 \land e^4\,.
\label{5D GR}
\end{equation}
While Eq.~\eqref{5D GB} is the 5-dimensional extension of the Gauss--Bonnet scalar, Eq.~\eqref{5D GR} represents the 5-dimensional expression of the Ricci scalar. Note that the former is not a topological term in 5 dimensions; however, once considering a 4-dimensional space-time, it reduces to
\begin{equation}
\epsilon_{abcd} R^{ab} \land  R^{cd} = 24 \frac{\ddot{a}\dot{a}^2}{a^3}\,,
\label{G4}
\end{equation}
which is nothing but $\G^{(5)}$ in the limit $b = 0$. When Eq.~\eqref{G4} is multiplied by the determinant of tetrad fields $|e|$, one gets the surface term
\begin{equation}
|e| \G^{(4)} = \frac{d}{dt}\left( 8 \dot{a}^3 \right).
\end{equation}
Replacing  Eqs.~\eqref{5D GB} and~\eqref{5D GR} into Eq.~\eqref{startact1}, the action can be written as:
\begin{equation}
S = \kappa \int \left\{\alpha_0 - \alpha_1 \left[6 \left(\frac{\ddot{a}}{a} + \frac{\dot{a}^2}{a^2} \right) + 2 \frac{\ddot{b}}{b} + 6 \frac{\dot{a}}{a} \frac{\dot{b}}{b}\right] + \alpha_2 \left[ 24\frac{\dot{a}^2 \ddot{a}}{a^3} + 48 \frac{\ddot{a} \dot{a}}{a^2} \frac{\dot{b}}{b} + 24 \frac{\dot{a}^2}{a^2} \frac{\ddot{b}}{b} + 24 \frac{\dot{a}^3}{a^3} \frac{\dot{b}}{b} \right]  \right\}e^0 \land e^1 \land e^2 \land e^3 \land e^4\,.
\end{equation}
By means of the relation in Eq.~\eqref{relazione prodotto esterno}, the above action takes the form:
\begin{eqnarray}
S = \kappa \int|e| \left\{\alpha_0 - \alpha_1 \left[6 \left(\frac{\ddot{a}}{a} + \frac{\dot{a}^2}{a^2} \right) + 2 \frac{\ddot{b}}{b} + 6 \frac{\dot{a}}{a} \frac{\dot{b}}{b}\right] + \alpha_2 \left[ 24\frac{\dot{a}^2 \ddot{a}}{a^3} + 48 \frac{\ddot{a} \dot{a}}{a^2} \frac{\dot{b}}{b} + 24 \frac{\dot{a}^2}{a^2} \frac{\ddot{b}}{b} + 24 \frac{\dot{a}^3}{a^3} \frac{\dot{b}}{b} \right]  \right\} d^5 x  \nonumber
\end{eqnarray}
\begin{equation}
= \int a^3b \left\{\alpha_0 + \alpha_1 \R^{(5)} + \alpha_2 \G^{(5)} \right\}d^5x\,.
\label{azione lovcosmo}
\end{equation}
After integrating by parts the terms containing second derivatives, the point-like cosmological Lagrangian turns out to be:
\begin{equation}
\Lagr = \alpha_0 a^3b +6 \alpha_1 \left(ab \dot{a}^2 + a^2 \dot{a} \dot{b} \right) - 8 \alpha_2 \dot{a}^3 \dot{b}\,.
\label{lagrangiana5dcosmo}
\end{equation}
The Euler--Lagrange equations and the energy condition are given by
\begin{eqnarray}
&& \frac{d}{dt} \frac{\partial \Lagr}{\partial \dot{a}} = \frac{\partial \Lagr}{\partial a} \;\;\; \to \;\;\; 4 \alpha_1 a ( \dot{a} \dot{b} + b \ddot{a}) - a^2 (\alpha_0 b - 2 \alpha_1 \ddot{b}) + 2 \dot{a} (\alpha_1 b \dot{a} - 8 \alpha_2 \dot{b} \ddot{a}- 4 \alpha_2 \dot{a} \ddot{b}) = 0\,, \nonumber
\\
&&\frac{d}{dt} \frac{\partial \Lagr}{\partial \dot{b}} = \frac{\partial \Lagr}{\partial b} \;\;\; \to \;\;\;  \alpha_0 a^3 - 6 \alpha_1 a \dot{a}^2 - 6 \alpha_1 a^2 \ddot{a} + 24 \alpha_2 \dot{a}^2 \ddot{a} = 0\,, \nonumber
\\
&& \dot{a} \frac{\partial \Lagr}{\partial \dot{a}} + \dot{b} \frac{\partial \Lagr}{\partial \dot{b}} - \Lagr = 0 \;\;\; \to \;\;\; \alpha_0 a^3 b - 6 \alpha_1 a b \dot{a}^2 - 6 \alpha_1 a^2  \dot{a} \dot{b}  + 24 \alpha_2 \dot{a}^3 \dot{b} = 0\,.
\label{EL5Dcosmo}
\end{eqnarray} 
In the limit $a(t) = b(t)$, the above system reduces to two differential equations, as the Euler--Lagrange equations with respect to $a(t)$ and $b(t)$ turn out to be equivalent as expected. Notice that the above Euler--Lagrange equations are of the fourth order and this can cause problems related to instability. This is mainly due to the fact that the Gauss--Bonnet invariant introduces higher derivatives with respect to the metric, so that the Lagrangian cannot be recast in a canonical form, with the result that the corresponding Hamiltonian might be linearly unstable. However, as pointed out in Refs.~\cite{Crisostomo:2000bb, Zwiebach:1985uq, Zumino:1985dp}, the action in Eq.~\eqref{startact1} leads to a ghost-free non-trivial gravitational interaction. More precisely, by expanding the action around the flat space-time, quadratic terms in the gravitational field combine to a total derivative and integrate to zero. As a consequence, Eq.~\eqref{startact1} introduces no propagator corrections. These considerations applies even when the higher-dimensional extension of Eq.~\eqref{startact1} is considered, as the action turns out to be ghost-free for any $d>3$. Generally, instability problems might occur whenever the Gauss--Bonnet scalar is considered, along with the scalar curvature, in the starting action. For instance, as pointed out in Ref.~\cite{DeFelice:2009ak}, $f(R,\G)$ theories could suffer the presence of ghost even at the level of cosmological perturbations, where superluminal modes can arise. This shortcoming was addressed \emph{e.g.} in Refs.~\cite{Astashenok:2015haa, Nojiri:2018ouv}, where the authors use Lagrange multipliers formalism to provide a remedy to ghost instability. Such a formalism leads to the elimination of the ghost degrees of freedom in both $f(\G)$ and $f(R,\G)$ gravity theories, and thus the resulting models can in principle produce ghost free primordial curvature perturbations. A particular solution of Eq.~\eqref{EL5Dcosmo} for the scale factors is:
\begin{equation}
a(t) = a_0 \exp\left\{\pm \sqrt{\frac{\alpha_0}{\pm 2 \sqrt{9
  \alpha_1^2-6 \alpha_0 \alpha_2}+6 \alpha_1}} \,  t \right\}\,, \;\;\;\;\;\; b(t) = b_0 \exp\left\{\pm \sqrt{\frac{\alpha_0}{\pm 2 \sqrt{9 \alpha_1^2-6 \alpha_0 \alpha_2}+6 \alpha_1}} \, t \right\}.
\label{SFLV}
\end{equation}
with $a_0$ and $b_0$ being arbitrary constants. Eq.~\eqref{SFLV} admits a constant Hubble horizon $r_H$ of the form
\begin{equation}
    r_H = \sqrt{\frac{6 \alpha_1 + 2 \sqrt{9 \alpha_1^2 - 6 \alpha_0 \alpha_2}}{\alpha_0}} \nonumber
\end{equation}
Notice that the only exponential solution admitted by the equations of motion, imposes the relation $a(t) \sim b(t)$. The limit $\alpha_0 = 0$, that is the sum between the scalar curvature and the Gauss--Bonnet term, provides the two further solutions:
\begin{eqnarray}
&& a(t) \sim b(t) \sim \text{Const.}
\\
&& a(t) = a_0 e^{\pm \sqrt{\frac{\alpha_1}{2 \alpha_2}}\, t} \sim b(t)
\label{GB5D}
\end{eqnarray}
The former can be obtained as the limit of Eq.~\eqref{SFLV} after taking the positive sign inside the square root. On the contrary, the latter cannot be recovered and must be computed by assuming a vanishing cosmological constant from the beginning. This is due to the fact that the solution with negative sign in Eq.~\eqref{SFLV} for $\alpha_0 = 0$ turns out to be indeterminate. Five-dimensional GR can be obtained by setting $\alpha_2 = 0$ and the field equations provide
\begin{equation}
a(t) = a_0  e^{\pm \sqrt{\frac{\alpha_0}{12 \alpha_1}}\, t} \sim b(t).
\label{GR5D}
\end{equation}
The similarity between Eqs.~\eqref{GB5D} and~\eqref{GR5D} suggests that in the former case the Gauss--Bonnet scalar plays the role of an effective cosmological constant. When the contribution of the scalar curvature vanishes, namely when $\alpha_1 = 0$, the Euler--Lagrange equations yield
\begin{equation}
a(t) = a_0 e^{\pm \left(\frac{\alpha_0}{-24\alpha_2}\right)^{1/4} \, t}\,\, \sim b(t).
\end{equation} 
According to Eqs.~\eqref{AdS2D-1} and~\eqref{CS5}, to get the AdS invariant Chern--Simons 5-dimensional Lagrangian we must set
\begin{equation}
\alpha_0 = \frac{1}{5 l^4}\,, \;\;\;\;\;\;\;\; \alpha_1 = \frac{2}{3 l^2}\,, \;\;\;\;\;\;\;\; \alpha_2 = 1\,,
\label{RelCS}
\end{equation}
so that the scale factors~\eqref{SFLV} become:
\begin{equation}
a(t) = a_0 \exp\left\{\pm \frac{1}{l} \sqrt{\frac{1}{6}\left(1\pm \sqrt{ \frac{7}{10}}\right)} \,\, t\right\}\,, \;\;\;\;\;\;\; b(t) = b_0 \exp\left\{\pm \frac{1}{l} \sqrt{\frac{1}{6}\left(1 \pm \sqrt{ \frac{7}{10}}\right)} \,\, t\right\} \,.
\label{CSa(t)}
\end{equation}
Note that the 3 free parameters occurring in the definition of the general Lovelock Lagrangian, are not constrained by the field equations solution and can be arbitrarily chosen. Thus both an exponential expansion and a bouncing evolution are admitted, depending on the values of the constants $\alpha_i$. On the other hand, imposing the relations in Eq.~\eqref{RelCS}, the cosmological solution in Eq.~\eqref{CSa(t)} suggests that only an accelerating expansion (or contraction) is allowed by the $AdS_5$ Chern--Simons cosmology.

\subsection{Generalization to $d+1$ Dimensions}

Let us consider now the extension of Lovelock gravity to $d+1$ dimensions, where, for convenience, we perform computations in spherical coordinates. In light of the result obtained in the previous Section, we set $b(t) = a(t)$ from the beginning, and we restore the spatial curvature $k$. The assumption of a unique scale factor $a(t)$ is needed to obtain suitable field equations, capable of providing analytic solutions in higher dimensions. The line element therefore takes the form
\begin{equation}
ds^2 = dt^2 - a(t)^2 \left[\frac{ dr^2}{1 - k r^2}  + r^2 d\Omega_{d-1}^2\right],
\end{equation}
where the $d-1$ sphere $d\Omega_{d-1}$ is defined through the space-time coordinates $x^\mu = (t,r,\theta_1,\theta_2,..., \theta_{d-2}, \theta_{d-1})$ as
\begin{equation}
d\Omega_{d-1}^2 = r^2 \left[d \theta_1^2 + \sum_{i=2}^{d-1} \prod_{j=1}^{i-1} \sin^2 \theta_j d\theta_i^2 \right].    \nonumber 
\end{equation}
We only consider the first 3 terms in the Lovelock $d+1$ dimensional Lagrangian, neglecting all those terms arising in more than 5 dimensions. This means that the coupling constants $\alpha_i$, with $i \ge 3$, will be set to zero. Consequently, the only non-vanishing terms in the action~\eqref{d-dim lovelock} are the $d+1$ dimensional scalar curvature $\R^{(d+1)}$ and the Gauss--Bonnet scalar $\G^{(d+1)}$. The action, therefore, is comprehended in the general Lovelock action in Eq.~\eqref{d-dim lovelock}, with a Lagrangian which generalizes the Lagrangian in Eq.~\eqref{lagrangiana5dcosmo}. Using the $d+1$ dimensional form of the Ricci scalar and of the Gauss--Bonnet term, namely
\begin{eqnarray}
&&\R^{(d+1)} = -d \left[2 \frac{\ddot{a}}{a} + (d-1) \left(\frac{\dot{a}^2 + k}{a^2} \right)\right],
\\
&&\G^{(d+1)} = d (d-1)(d-2) \left[(d-3) \left(\frac{\dot{a}^4}{a^4} + 2 k \frac{\dot{a}^2}{a^4} + \frac{k^2}{a^4}\right) + \frac{4}{3 a^3} \frac{d}{dt}\left( \dot{a}^3\right) + 4 k \frac{\ddot{a}}{a^3}\right],
\end{eqnarray}
and considering a starting action of the form
\begin{equation} 
S = \kappa \int |e| \left[\alpha_0 + \alpha_1 \R^{(d+1)} + \alpha_2 \G^{(d+1)}\right] d^{d+1} x\,,
\end{equation} 
the Lagrangian can be written as:
\begin{equation}
\Lagr = \frac{r^{d-1} a^{d-4} \prod_{j=2}^{d-1} (\sin \theta_{j-1})^{d-j}}{3 \sqrt{1 - k r^2}} \left\{3 a^2 [a^2 \alpha_0 + \alpha_1 d (d-1) (\dot{a}^2-k)]-\alpha_2 d (d-1)(d-2)(d-3) (\dot{a}^4 + 6 k \dot{a}^2 - 3 k^2)\right\}\,.
\label{lagrk}
\end{equation}
The corresponding equation of motion with respect to the scale factor and the energy condition read, respectively:
\begin{eqnarray}
&&-2 \alpha_1 (d-1) a^3 \ddot{a}-\alpha_1 (d-1)(d-2) a^2
   \left(\dot{a}^2+k\right)+\alpha_2 (d-1)(d-2)(d-3)(d-4)
   \left(\dot{a}^2+k\right)^2 \nonumber 
   \\
 &&  +4 \alpha_2 (d-1)(d-2)(d-3) a \ddot{a} \left(\dot{a}^2+k\right)+\alpha_0 a^4= 0\,, \nonumber
    \\
&&    \alpha_2 d \left(d^3-6 d^2+11 d-6\right) \left(\dot{a}^2+k\right)^2-\alpha_1 d (d-1) a^2 \left(\dot{a}^2+k\right)+\alpha_0 a^4 = 0\,.
\end{eqnarray}
Notice that, when setting $k=0$, the above system of differential equations reduces to that in Eq. \eqref{EL5Dcosmo} with $a(t) = b(t)$. The Gauss--Bonnet term, as confirmed by the above equations, does not contribute to the equations of motion in less than $3+1$ dimensions, where only the scalar curvature plays a role in the dynamics. Let us first consider the case $k \neq 0$, where no solutions occur if all the coupling constants are simultaneously non-zero. Setting $\alpha_2 = 0$, the theory reduces to the higher-dimensional GR with cosmological constant, in a spatially non-flat universe. The scale factor which solves the equations of motion, in this case turns out to be
\begin{equation}
a(t) = \pm \sqrt{-\frac{\alpha_1 k d(d-1)}{\alpha_0}} \sinh \left[\sqrt{\alpha_0}\left(c_1 + \frac{t}{\sqrt{\alpha_1 d(d-1)}} \right) \right]
\label{SFDS}
\end{equation}
which holds as long as $\alpha_0 \neq 0$. In this general case, the Hubble horizon sits at
\begin{equation}
r_H=\sqrt{\frac{\alpha_1  d(d-1)}{\alpha_0}} \tanh \left[\sqrt{\alpha_0}\left(c_1 + \frac{t}{\sqrt{\alpha_1 d(d-1)}} \right) \right] \nonumber
\end{equation}
When setting $\alpha_0 = 0$ from the beginning, the only solution is:
\begin{equation}
a(t) =  \sqrt{-k} \, t\,.
\label{SFGR}
\end{equation}
An other analytical solution occurs for $\alpha_1 = \alpha_0 = 0$, where the Euler--Lagrange equations and the energy condition yield the scale factor
\begin{equation}
a(t) = \sqrt{-k} \, t\,,
\label{SFGB}
\end{equation}
which is exactly the same as Eq.~\eqref{SFGR}. Finally, when $\alpha_1 \neq 0$, $\alpha_2 \neq 0 $, and $\alpha_0 = 0$, the only solution is: 
\begin{equation}
a(t) = \pm \sqrt{\frac{- \alpha_2 k (d-3) (d-2)}{\alpha_1}} \sinh \left[\sqrt{\alpha_1} \left(\frac{t}{\sqrt{\alpha_2 (d-3)(d-2)}}+c_1\right)\right],
\label{SFGBGR}
\end{equation}
with Hubble horizon
\begin{equation}
r_H = \sqrt{\frac{\alpha_2 (d-3) (d-2)}{\alpha_1}} \tanh \left[\sqrt{\alpha_1} \left(\frac{t}{\sqrt{\alpha_2 (d-3)(d-2)}}+c_1\right)\right]. \nonumber   
\end{equation}
Notice that Eq.~\eqref{SFGBGR} is equivalent to the cosmological solution provided in Ref.~\cite{Wheeler:1985nh}, but with different definitions of the integration parameters. Comparing Eq.~\eqref{SFGBGR} with Eq.~\eqref{SFDS}, we notice that the Gauss--Bonnet term can play the role of an effective cosmological constant. 

Let us now focus on a spatially flat FLRW space-time in higher dimensions. When $k = 0$, the Lagrangian~\eqref{lagrk} reduces to:
\begin{equation}
\Lagr = \frac{r^{d-1} a^{d-4} \prod_{j=2}^{d-1} (\sin \theta_{j-1})^{d-j}}{3} \left\{3 a^2 [a^2 \alpha_0 + \alpha_1 d (d-1) \dot{a}^2]-\alpha_2 d (d-1)(d-2)(d-3) \dot{a}^4 \right\}\,.
\label{lagrcosmoLL}
\end{equation}
The general solution of the related Euler--Lagrange equations can be analytically found only for exponential scale factor of the form
\begin{equation}
a(t)= a_0 \exp\left\{\pm \sqrt{\frac{2 \alpha_0}{\pm \sqrt{(d-1) d \left[\alpha_1^2 (d-1)
   d-4 \alpha_0 \alpha_2 (d-3) (d-2)\right]}+\alpha_1 d(d-1)}} \, t\right\}.
   \label{gensold+1}
\end{equation}
Notice that in the 5-dimensional limit, the scale factor reduces to~\eqref{SFLV}, namely
\begin{equation}
a^{(5)}(t)= a_0 \exp\left\{\pm \sqrt{\frac{\alpha_0}{\pm 2 \sqrt{9
  \alpha_1^2- 6 \alpha_0 \alpha_2}+6 \alpha_1}} \, \, t\right\}.
\end{equation}
Let us now separately analyze the subcases not covered by the solution~\eqref{gensold+1}, namely those cases in which (at least) one of the constants $\alpha_i$ vanishes; in the limit $\alpha_2 = 0$, Lagrangian~\eqref{lagrcosmoLL} turns into the high-dimensional Einstein--Hilbert Lagrangian with cosmological constant, providing the well known Einstein--de Sitter solution
\begin{equation}
a(t) = a_0  e^{\pm \sqrt{\frac{\alpha_0}{\alpha_1 d(d-1)}}\, t}\,,
\end{equation}
which is the spatially flat limit of Eq.~\eqref{SFDS}.

The case $\alpha_1 = 0$ (analyzed in depth in Ref.~\cite{Bajardi:2020osh}) yields the following non-trivial solution:
\begin{equation}
a(t) = a_0 \exp\left\{\pm\left[-\frac{\alpha_0}{\alpha_2} \left(\frac{1}{d(d-1)(d-2)(d-3)} \right)\right]^{1/4} \, t\right\}.
\end{equation} 
Finally, by setting $\alpha_0 = 0$, a de Sitter-like scale factor of the form 
\begin{equation}
a(t) = a_0 e^{\pm \sqrt{\frac{\alpha_1}{\alpha_2 \left(d-2\right)(d-3)}}\, t}\,,
\label{SFGBGRk0}
\end{equation}
solves the field equations. This subcase cannot be directly recovered from Eq.~\eqref{gensold+1}, whose $\alpha_0 = 0$ limit provides an indeterminate scale factor. Also here, the values assumed by the coupling constants can determine either an exponential expansion or an oscillating solution. Moreover, Eq.~\eqref{SFGBGRk0} is the $k = 0$ limit of Eq.~\eqref{SFGBGR}.

Notice that no solution occurs in vacuum when two coupling constants are simultaneously null; therefore, the cases analyzed here are all the possible subcases that can be obtained from Lagrangian in Eq.~\eqref{lagrcosmoLL}. This can be also inferred from Eqs.~\eqref{SFGR} and~\eqref{SFGB}, whose $k = 0$ limits provide $a(t) = 0$.

For a detailed discussion regarding Lovelock cosmology see \emph{e.g.} Ref.~\cite{Deruelle:1989fj, Deruelle:1987ux}, where the authors provide a systematic study of cosmological solutions including maximally symmetric space-times. 

\section{Lovelock and $AdS_5$ Chern--Simons Gravity in a Spherically Symmetric Background}
\label{CSSS}

Here we study a spherically symmetric background in Lovelock gravity and find out exact solutions. Though some of them are already present in the literature (such as few related to higher-dimensional Lovelock action~\cite{Myers:1988ze, Aros:2000ij, Myers:1989kt}), some others are new, as well as the analysis of horizon and black hole entropies. Interestingly, the features of black hole entropies in higher-dimensional Lovelock gravity have been considered in Ref.~\cite{Jacobson:1993vj}, where the authors found a link between entropies and Noether charges. Moreover, in Ref.~\cite{Jacobson:1993xs}, T.~Jacobson and R.~Myers obtained a general formula for the entropy of black holes in Lovelock gravity, which generally differ from the standard Bekenstein--Hawking one. However, additional contributions to the standard Bekenstein--Hawking entropy become relevant when higher curvature invariants are included into the starting Lovelock action. Here, such as the cosmological case, we only consider the first three Lovelock parameters, so that the action is made of a sum of the scalar curvature, the Gauss--Bonnet term and the cosmological term. As a consequence, the entropy is proportional to the black hole area, as standard, even in higher than five dimensions.

As we did in the previous Section, the 5-dimensional case is firstly analyzed. Subsequently, we generalize the treatment to $d+1$ dimensions, where all the coupling constants $\alpha_{i}$, with $i>2$, are neglected. In this way, only the higher-dimensional generalization of the scalar curvature and of the Gauss--Bonnet term are considered into the action. This \emph{ansatz} allows to solve the field equations analytically. For the  reasons mentioned above, we have to pay main attention to the 5-dimensional case, which needs to be treated separately. For particular combinations of the coupling constants, the 5-dimensional Lovelock Lagrangian reduces to the Chern--Simons Lagrangian, invariant under the local AdS group. On the other hand, while in $d+1$ dimensions analytical solutions can be found only under the assumption $P(r) = Q(r)^{-1}$ for the metric coefficients, in 5 dimensions, exact solutions occur without adopting any extra \emph{ansatz}. Let us then consider the metric:
\begin{equation}
ds^2 = P(r)^2 dt^2 - Q(r)^2 dr^2 - r^2 d\theta^2 - r^2 \sin^2(\theta) d\phi^2 - r^2 \sin^2 \theta \sin^2 \phi \, d \psi^2\,, 
\label{INT5}
\end{equation}
where $d\theta^2 + \sin^2(\theta) d\phi^2 + \sin^2 \theta \sin^2 \phi d \psi^2 \equiv d\Omega_{3}$ is the 3-sphere. By choosing the following set of basis
\begin{equation}
e^0 =P(r) dt\,, \;\;\; e^1 = -Q(r) dr\,, \;\;\; e^2 = -r d\theta\,, \;\;\; e^3 = -r \sin(\theta) d\phi\,, \;\;\; e^4 = -r \sin(\theta) \sin(\phi) d \psi\,,
\end{equation}
the same computations as in the previous Section yield the expressions of the Gauss--Bonnet term and of the Ricci scalar in the spherically symmetric background
\begin{eqnarray}
\G^{(5)} = &-&\frac{24 P''(r)}{r^2 P(r) Q(r)^2}+\frac{24 P''(r)}{r^2 P(r) Q(r)^4}+\frac{24 P'(r)
   Q'(r)}{r^2 P(r) Q(r)^3}-\frac{72 P'(r) Q'(r)}{r^2 P(r) Q(r)^5}-\frac{24 P'(r)}{r^3
   P(r) Q(r)^2}  \nonumber
   \\
   &+&\frac{24 P'(r)}{r^3 P(r) Q(r)^4}+\frac{24 Q'(r)}{r^3 Q(r)^3}-\frac{24
   Q'(r)}{r^3 Q(r)^5}\,,
\label{Gaussbonnet}
\end{eqnarray}
\begin{equation}
\R^{(5)} = \frac{2 P''(r)}{P(r) Q(r)^2}-\frac{2 P'(r) Q'(r)}{P(r) Q(r)^3}+\frac{6 P'(r)}{r P(r)
   Q(r)^2}-\frac{6 Q'(r)}{r Q(r)^3}+\frac{6}{r^2 Q(r)^2}-\frac{6}{r^2}\,. \;\;\;\;\;\;\;\;\;\;\;\;\;\;\;\;\;\;\;\;
   \label{Ricci}
\end{equation}
The Lovelock Lagrangian therefore can be written as:
\begin{eqnarray}
&& \Lagr^{(5)} =  \sin^2 \theta \sin \phi \left\{\alpha_0 r^3 P(r) Q(r) - \alpha_1 \left(\frac{6 r^2 P(r) Q'(r)}{Q(r)^2}+6 r P(r) Q(r)-\frac{6 r
   P(r)}{Q(r)}\right) \right. \nonumber \\ 
   && \qquad \qquad \qquad \qquad + \left. \alpha_2 \left(\frac{24 P(r) Q'(r)}{Q(r)^2}-\frac{24 P(r)
   Q'(r)}{Q(r)^4}\right) \right\}\,.
   \label{lovelock point like}
\end{eqnarray}
Since the values of the coupling constants play a fundamental role in the treatment, in the next Subsection we will analyze the contribution of each term separately, to finally provide the spherically symmetric solutions of the field equations for $AdS_5$ Chern--Simons and Lovelock gravity. 

\subsubsection{5-dimensional Einstein gravity}

Let us start by setting $\alpha_2=0$, so that the Lagrangian reduces to the $4+1$ dimensional Einstein--Hilbert Lagrangian with cosmological constant $\alpha_0$, namely
\begin{equation}
\Lagr = \sin^2 \theta \sin \phi \left\{\alpha_0 r^3 P(r) Q(r) - \alpha_1 \left(\frac{6 r^2 P(r) Q'(r)}{Q(r)^2}+6 r P(r) Q(r)-\frac{6 r
   P(r)}{Q(r)}\right)\,\right\}.
\end{equation}
The following Euler--Lagrange equations 
\begin{equation}
\begin{split}
&\frac{d}{dr} \frac{\partial \Lagr}{\partial P'(r)} = \frac{\partial \Lagr}{\partial P(r)} \;\; \to \;\; \alpha_0 r^3 Q(r) = 6 r \alpha_1 \left(\frac{r Q'(r)}{Q(r)^2}+ Q(r)-\frac{1}{Q(r)}\right),
\\
& \frac{d}{dr} \frac{\partial \Lagr}{\partial Q'(r)} = \frac{\partial \Lagr}{\partial Q(r)} \;\; \to \;\; \alpha_0 r^3 P(r) = 6 r \alpha_1 \left(P(r)- \frac{P(r)}{Q(r)^2} -  \frac{r P'(r)}{Q(r)^2}\right), \nonumber
\end{split}
\end{equation}
provide the solution
\begin{equation}
P(r)^2 = 1 + \frac{c_1}{r^2} - \frac{\alpha_0}{12 \alpha_1} r^2\,, \;\;\;\;\;\;\;\;\;\;\;\;\;\;\;\;\;\;  Q(r)^2 = \displaystyle \frac{1}{1 + \frac{c_1}{r^2} - \frac{\alpha_0}{12 \alpha_1} r^2}\,,
\end{equation}
where $c_1$ is an integration constant. The above solution is the 5-dimensional extension of an Einstein--de Sitter space-time~\cite{Ahmed:2014eea}. Notice that $P(r)$ vanishes when
\begin{equation}
r =  r_H = \sqrt{\frac{2 \sqrt{3 \alpha_1 (3 \alpha_1 +\alpha_0 c_1)}}{\alpha_0}+\frac{6 \alpha_1}{\alpha_0}}\,.
\end{equation}
Moreover, setting $\alpha_0 = 0$, the horizon turns out to be $r_H =  \sqrt{- c_1}$\,. This means that the quantity $-c_1$ can be intended as a mass term in 5 dimensions. Therefore, the Bekenstein--Hawking entropy $\mathcal{S}$, in the Einstein--de Sitter 5-dimensional space-time, is
\begin{equation}
    \mathcal{S} = \frac{1}{2} \pi^2 r_H^3 \sim M^{\frac{3}{4}}\,,
\end{equation}
which is exactly the same dependence of the entropy exhibited by the entropy of a 4-dimensional conformal field theory (CFT)~\cite{Myung:2001ab, Aharony:1998tt}, because of the AdS/CFT correspondence. The same result is not provided by the 5-dimensional Einstein gravity with $\alpha_0 = 0$, where the entropy behaves like
\begin{equation}
    \mathcal{S} \sim M^{\frac{3}{2}}.
\end{equation}
An accurate analysis is reported in Sec.~\ref{GEND}, where the discussion is extended to $d+1$ dimensions. 

\subsubsection{Pure Gauss--Bonnet gravity}

Let us now consider the case $\alpha_1 = 0$, in which the Lagrangian is only made by the Gauss--Bonnet term and the cosmological constant. The only terms which survive in the Lagrangian after integrating the second derivatives are:
\begin{equation}
\Lagr = \sin^2 \theta \sin \phi \left\{\alpha_0 P(r) Q(r) r^3 + \alpha_2 \left(\frac{24 P(r) Q'(r)}{Q(r)^2}-\frac{24 P(r) Q'(r)}{Q(r)^4}\right)\right\}  \,,
\label{Lagr gauss-bonnet}
\end{equation}
and the Euler--Lagrange equations read
\begin{equation}
\begin{split}
&\frac{d}{dr} \frac{\partial \Lagr}{\partial P'(r)} = \frac{\partial \Lagr}{\partial P(r)} \;\; \to \;\; -24 \alpha_2 \left(Q(r)^2-1\right) Q'(r)= \alpha_0 r^3 Q(r)^5\,,
\\
& \frac{d}{dr} \frac{\partial \Lagr}{\partial Q'(r)} = \frac{\partial \Lagr}{\partial Q(r)} \;\; \to \;\; 24 \alpha_2\left(Q(r)^2-1\right) P'(r)= \alpha_0 r^3 P(r) Q(r)^4\,.
\end{split} 
\label{ELGBo}
\end{equation}
There are two classes of solutions coming from the above equations. The first arises by imposing $P(r) = 1/Q(r)$ between the metric components $P(r)$ and $Q(r)$. It reads: 
\begin{eqnarray}
&&P(r)^2 =1\pm \sqrt{1+ c_1 - \frac{\alpha_0}{24 \alpha_2} r^4}\,,    \label{BirkGB}
\\
&& Q(r) = \frac{1}{P(r)}\,,
   \label{BirkGB1}
\end{eqnarray} 
and the horizon, only present for the ``minus'' branch of the solution, sits at 
\begin{equation}
\displaystyle r_H = \left( \frac{24 \alpha_2 \, c_1}{\alpha_0}\right)^{1/4}\,,
\label{HorizonGB}
\end{equation} 
where $c_1$ is an integration constant. As we can see from Eqs.~\eqref{BirkGB} and~\eqref{BirkGB1}, by setting $\alpha_0 = 0$ only trivial solutions occur. This means that, under the assumption $P(r) = 1/Q(r)$, the only non-trivial contribution in 5 dimensions is provided by the cosmological constant. Moreover, by using the horizon in Eq.~\eqref{HorizonGB}, the entropy turns out to be
\begin{equation}
    \mathcal{S} = \frac{1}{2} \pi^2  \left( \frac{24 \alpha_2}{\alpha_0}\right)^{3/4} M^{3/4}\,,
\end{equation} 
and the AdS/CFT correspondence is recovered. 

An other line element, which solves Eqs.~\eqref{ELGBo} in 5 dimensions, can be found without imposing any relation between $P(r)$ and $Q(r)$. In this case, the solution is:
\begin{eqnarray}
&& P^2(r) = P_0^2 \sqrt{48 \alpha_2 \left(2 c_2+1\right) \mp 4 \sqrt{6} \sqrt{\alpha_2
   \left(24 \alpha_2 +96 \alpha_2  c_2-\alpha_0
   r^4\right)}-\alpha_0 r^4}\,,   \label{SOLPURGBP}
   \\
   && Q^2(r) = \frac{2 \left(12 \alpha_2 \pm \sqrt{6} \sqrt{\alpha_2 \left(24
  \alpha_2+96 \alpha_2 c_2-\alpha_0
   r^4\right)}\right)}{\alpha_0 r^4-96 \alpha_2 c_2}\,,
   \label{SOLPURGBQ}
\end{eqnarray}
where $P_0$ and $c_2$ are integration constants. The horizon can be found by setting $P(r) = 0$, from which four mathematical solutions follow, but the only one with physical meaning is
\begin{equation}
r_H = 2 \left( \frac{6 \alpha_2 \, c_2}{\alpha_0}\right)^{1/4}\,.
\end{equation}
Notice that $Q(r_H)$ diverges, though the \emph{ansatz} $P(r) = 1/Q(r)$ is not imposed from the beginning. Also, after redefining the constant $c_2$, this horizon turns out to be the same as Eq.~\eqref{HorizonGB}. This is expected since the solution in Eqs.~\eqref{BirkGB} and~\eqref{BirkGB1}, coming from the imposition $P(r)=Q(r)^{-1}$, is a particular subcase of Eq.~\eqref{SOLPURGBP}.

As an example, a more suitable form can be obtained by setting $c_2 = -1/4$, where Eqs.~\eqref{SOLPURGBP} and~\eqref{SOLPURGBQ} reduce to:
\begin{eqnarray}
&& P^2(r) = P_0^2 \sqrt{24 \alpha_2 \mp 4 \sqrt{6} \sqrt{-\alpha_2 \alpha_0 r^4}-\alpha_0 r^4}\,,
   \\
   && Q^2(r) = \frac{24 \alpha_2 \pm 2 \sqrt{6} \sqrt{-\alpha_2 \alpha_0 r^4}}{\alpha_0 r^4+24 \alpha_2 }\,.
\end{eqnarray}
In any case, the metric turns out to be trivially constant when the cosmological constant $\alpha_0$ is not considered. Moreover, in the limit $r \to \infty$, the asymptotic flatness is not recovered for $P(r)$. 
However, as we will point out in Sec.~\ref{GEND}, in higher dimensions, the flatness for large radius and the presence of the horizon can be obtained for some combinations of the coupling constants.

\subsubsection{Lovelock and $AdS_5$ Chern--Simons Gravity}

To conclude the search for exact solutions in the 5-dimensional space-time, we find Euler--Lagrange equations solutions in the most general case, where no coupling constants are neglected in the action~\eqref{lovelock point like}. The equations of motion are:
\begin{equation}
\begin{split}
&\frac{d}{dr} \frac{\partial \Lagr}{\partial P'(r)} = \frac{\partial \Lagr}{\partial P(r)} \;\; \to \;\;  Q(r) \left(\alpha_0 r^3-6 \alpha_1 r\right)-\frac{6 \left(\alpha_1 r^2-4 \alpha_2 \right) Q'(r)}{Q(r)^2}+\frac{6 \alpha_1 r}{Q(r)}-\frac{24 \alpha_2 Q'(r)}{Q(r)^4} = 0\,,
\\
& \frac{d}{dr} \frac{\partial \Lagr}{\partial Q'(r)} = \frac{\partial \Lagr}{\partial Q(r)} \;\; \to \;\; 6 P'(r) \left\{Q(r)^2 \left[-\alpha_1 r^2+4 \alpha_2\right]-4 \alpha_2 \right\} - r P(r) Q(r)^2 \left[  Q(r)^2 \left( \alpha_0 r^2-6 \alpha_1 \right) +6 \alpha_1 \right] = 0\,.
\end{split}
\label{EOMLOV}
\end{equation}
Here, the imposition $P(r) = 1/Q(r)$ is not required to solve the equations of motion. As a matter of fact, the general solution of the Euler--Lagrange equations is:
\begin{eqnarray}
P(r)^2=P_0^2 && \sqrt{-4 c_1+\alpha_0 r^4-12 \alpha_1 r^2}
   \left[\frac{3 \sqrt{r^4 u w+2 u x}+\sqrt{3} \left(\alpha_0 x+r^2 w (3
   \alpha_1-z)\right)}{-6 \alpha_1+\alpha_0 r^2+2
   z}\right]^{y/2} \nonumber
   \\
   &&\times \left[\frac{6 \alpha_1-\alpha_0 r^2+2 z}{3
   \sqrt{r^4 v w+2 v x}+\sqrt{3} \left(r^2 w (z+3 \alpha_1)+\alpha_0
   x\right)}\right]^{s/2}\,,
   \label{P11*}
\end{eqnarray}
\begin{equation}
Q(r)^2 = \frac{-3 \alpha_1 r^2+12 \alpha_2 \pm \sqrt{3} \sqrt{8 c_1 \alpha_2 -2 \alpha_0 \alpha_2 r^4+3 \alpha_1^2 r^4+48 \alpha_2^2}}{\frac{\alpha_0}{2} r^4-6 \alpha_1 r^2 -2
   c_1}\,,
   \label{Q11*}
\end{equation}
where the constants $s,u,v,w,x,y,z$, arising in $P(r)^2$, are defined in Appendix~\ref{APPB}. By setting $P(r) = 0$,  the following  horizons occur, namely:
\begin{eqnarray}
r_H &=& \sqrt{\pm \frac{2 \sqrt{c_1 \alpha_0 + 9 \alpha_1^2}}{\alpha_0}+6 \frac{\alpha_1}{\alpha_0}}\,, \nonumber
\\  \nonumber \\ \nonumber 
r_H &=& \left[\pm\frac{\sqrt{3} \sqrt{-6 u^2 w x+18 \alpha_1^2 u w^2 x+2 u w^2 x z^2-12
   \alpha_1 u w^2 x z+\alpha_0^2 u w x^2}}{3 u w-9 \alpha_1^2
   w^2 - w^2 z^2 +6 \alpha_1 w^2 z}\right. \nonumber
   \\
   && \left.-\frac{\alpha_0 w x z}{3
   u w-9 \alpha_1^2 w^2-w^2 z^2+6 \alpha_1 w^2 z}+\frac{3
   \alpha_0 \alpha_1 w x}{3 u w-9 \alpha_1^2 w^2-w^2
   z^2+6 \alpha_1 w^2 z}\right]^{1/2}\,, \nonumber
   \\ \nonumber \\ 
   r_H &=& \sqrt{\frac{-2z+6\alpha_1}{\alpha_0}}\,.
   \label{HorizonsLOV}
\end{eqnarray}
They can be obtained by imposing the first, second and third term in Eq.~\eqref{P11*} to be equal to zero, respectively.
Notice that $Q(r)$ diverges only for the first couple of horizons, that is:
\begin{equation}
Q(r_H) \to \infty \quad \text{where} \quad r_H = \sqrt{\pm \frac{2 \sqrt{c_1 \alpha_0 + 9 \alpha_1^2}}{\alpha_0}+6 \frac{\alpha_1}{\alpha_0}}\,,
\end{equation}
with a Bekenstein--Hawking entropy of the form
\begin{equation}
\mathcal{S} = \frac{1}{2} \pi^2 \left(\pm \frac{2 \sqrt{M \alpha_0 + 9 \alpha_1^2}}{\alpha_0}+6 \frac{\alpha_1}{\alpha_0}\right)^{3/2},
\end{equation}
which scales as that of a CFT. It is worth discussing the behavior of the general solution in Eq.~\eqref{P11*} for some particular values of the coupling constants $\alpha_i$. By setting $\alpha_2 = 0$, the standard 5-dimensional Einstein gravity with cosmological constant is recovered, while in the limit $\alpha_1=0$ Eqs.~\eqref{P11*} and~\eqref{Q11*} reduce to Eqs.~\eqref{SOLPURGBP} and~\eqref{SOLPURGBQ} (with a proper rescaling of the integration constant $c_1$). By neglecting the cosmological constant (\emph{i.e.} setting $\alpha_0 = 0$), the result can be simplified and two different solutions can be analytically found. They read
\begin{equation}
P_-(r)^2 = P_0^2 \left(2 c_1+ \alpha_1 r^2\right) \sqrt{\frac{\sqrt{16 \alpha_2 (\alpha_2+c_1)+\alpha_1^2 r^4}+\alpha_1 r^2}{8 \alpha_2^2+2 \alpha_2 \left(\sqrt{16 \alpha_2 (\alpha_2+c_1)+\alpha_1^2 r^4}+4 c_1\right)+c_1 \left(\sqrt{16 \alpha_2
   (\alpha_2+c_1)+\alpha_1^2 r^4}-\alpha_1 r^2\right)}}\,,
   \label{PLOV}
\end{equation}
\begin{equation}
 P_+(r)^2 = P_0^2 \sqrt{\frac{8 \alpha_2^2+2 \alpha_2 \left(\sqrt{16 \alpha_2 (\alpha_2 + c_1)+\alpha_1^2 r^4}+4 c_1\right)+c_1 \left(\sqrt{16
   \alpha_2 (\alpha_2+c_1)+\alpha_1^2 r^4}-\alpha_1 r^2\right)}{\sqrt{16 \alpha_2 (\alpha_2+c_1)+\alpha_1^2 r^4}+\alpha_1 r^2}}\,,
\end{equation}
\begin{equation}
Q_{\pm}(r)^2 = \frac{-4 \alpha_2 \pm \sqrt{16 \alpha_2 (\alpha_2 +c_1)+\alpha_1^2 r^4}+\alpha_1 r^2}{4 c_1+2 \alpha_1 r^2}\,.
\label{Q22*}
\end{equation}
Both solutions share the same horizon, namely
\begin{equation}
    r_H = \sqrt{\frac{-2c_1}{\alpha_1}}\,,
\end{equation}
so that the corresponding Bekenstein--Hawking entropy is
\begin{equation}
\mathcal{S} =   \pi^2 \sqrt{\frac{2}{\alpha_1^3}} M^{\frac{3}{2}}\,,
\end{equation}
which behaves differently than that of a CFT, having neglected the cosmological constant $\alpha_0$.

In the former case, the asymptotic flatness cannot be recovered, independently of the value of the integration constant $P_0$. In the latter case, the flatness for large radius occurs by means of the choice $P_0^2 = (\alpha_2)^{-1/2}$. 

The 5-dimensional Chern--Simons gravity, invariant under the local AdS group, can be found as a limit of Eqs.~\eqref{PLOV} and~\eqref{Q22*}, imposing $\displaystyle \alpha_2 = 1$, $\displaystyle \alpha_1 = \frac{2}{3 l^2}$, $ \displaystyle \alpha_0 = \frac{1}{5 l^4}$. In such a case, the point-like spherically symmetric Lagrangian (up to the term $\sin^2 \theta \sin \phi$) reads:
\begin{equation}
 \Lagr^{(5-AdS)}_{CS} = \frac{\kappa}{l} \left[-\frac{2}{3l^2} \left(\frac{6 r^2 P(r) Q'(r)}{Q(r)^2}+6 r P(r) Q(r)-\frac{6 r
   P(r)}{Q(r)}\right)+ \left(\frac{24 P(r) Q'(r)}{Q(r)^2}-\frac{24 P(r)
   Q'(r)}{Q(r)^4}\right) + \frac{1}{5l^4} r^3 P(r) Q(r)  \right],
\label{lagr CS}
\end{equation}
and the Euler--Lagrange equations yield
\begin{eqnarray}
P(r)^2=P_0^2 && \sqrt{-4 c_1 +\frac{r^4}{5 l^4}-\frac{8 r^2}{l^2}}
   \left[\frac{\sqrt{3} \left(\frac{x}{5 l^4}-r^2 w
   \left(-\frac{2}{l^2}+z\right)\right)+3 \sqrt{u \left(r^4 w+2 x\right)}}{\frac{r^2}{5
   l^4}-\frac{4}{l^2}+2 z}\right]^{y/2} 
   \\
 &&   \times\left[\frac{-\frac{r^2}{5 l^4}+\frac{4}{l^2}+2
   z}{\sqrt{3} \left(\frac{x}{5 l^4}+r^2 w \left(z+\frac{2}{l^2}\right)\right)+3 \sqrt{v
   \left(r^4 w+2 x\right)}}\right]^{s/2}\,,\nonumber
\end{eqnarray}
\begin{equation}
Q(r)^2 = \frac{\sqrt{24 c_1 + \frac{14 r^4}{5 l^4}+144}-\frac{2
   r^2}{l^2}+12}{-2 c_1 +\frac{r^4}{10 l^4}-\frac{4 r^2}{l^2}}\,.
   \label{CSQ1}
\end{equation}
The form of $P(r)$ is  the same as in the case of Lovelock gravity, with different values of the coupling constants (see Appendix~\ref{APPB}). Also the horizons are formally the same as those in Eq.~\eqref{HorizonsLOV}. They read:
\begin{eqnarray}
r_H &=& l \, \sqrt{20 \pm 2 \sqrt{100 + 5 c_1}}\,, \nonumber
\\  \nonumber \\ \nonumber 
r_H &=& \left[\pm \frac{\sqrt{3} \sqrt{-150 l^8 u^2 w x+50 l^8 u w^2 x z^2-200 l^6 u w^2 x
   z+200 l^4 u w^2 x+u w x^2}}{5 \left(3 l^4 u w-l^4 w^2 z^2-4 l^2 w^2 z-4
   w^2\right)}\right. \nonumber
   \\
   &&\left. +\frac{2 w x}{5 l^2 \left(3 l^4 u w-l^4 w^2 z^2+4 l^2 w^2 z-4
   w^2\right)}-\frac{w x z}{5 \left(3 l^4 u w-l^4 w^2 z^2+4 l^2 w^2 z-4
   w^2\right)}\right]^{1/2}\,, \nonumber
   \\ \nonumber \\ 
   r_H &=& l^2 \sqrt{10} \sqrt{l^2z +2}\,.
   \label{HorizonsCS}
\end{eqnarray}
The limit of large radius does not provide a flat Minkowski space-time, since the cosmological constant cannot be assumed to vanish. However, it is worth noticing that, for large value of $l$, the pure Gauss--Bonnet gravity is restored, while for $l \ll 1$ we get an Einstein--de Sitter space-time.

As mentioned above, from Eqs.~\eqref{EOMLOV}, an other subclass of solutions occurs, constrained by the imposition $P(r) = 1/Q(r)$, reads as:
\begin{eqnarray}
&&P(r)^2 = 1-\frac{\alpha_1 r^2}{4
   \alpha_2} \pm \frac{\sqrt{3 r^{4} \left(3 \alpha_1^2-2 \alpha_0
   \alpha_2\right)+6 \alpha_2 c_1 }}{12 \alpha_2}\,,\nonumber
\\
&&Q(r) = 1/P(r)\,.
\label{Q1*}
\end{eqnarray}
This particular 5-dimensional solution has already been found and studied \emph{e.g.} in Refs.~\cite{Garraffo:2008hu} and~\cite{Anabalon:2009kq}. Specifically, in the latter reference, the authors consider 5-dimensional rotating Lovelock black holes and find out exact solutions. 

The two horizons, by means of an appropriate redefinition of the integration constant $c_1$, are exactly the same as the first couple of Eqs.~\eqref{HorizonsLOV}. Without the cosmological constant, solution \eqref{Q1*} takes the form
\begin{equation}
P(r)^2 = 1-\frac{\alpha_1 r^2}{4
   \alpha_2} \pm \frac{\sqrt{9 \alpha_1^2 r^{4}+6 \alpha_2 c_1 }}{12 \alpha_2}\,.
\label{Q2*}
\end{equation}
The Chern--Simons solution can be recovered as a particular limit of Eq.~\eqref{Q1*}, namely:
\begin{equation}
P(r)^2 = 1-\frac{r^2}{6 l^2}\pm  \sqrt{\frac{7 r^{4}}{360 l^4}+\frac{c_1}{24}}\,,
\end{equation}
with horizons sitting at
\begin{equation}
r_H = l \, \sqrt{20 \pm \sqrt{5 c_1 + 280}}\,.
\end{equation}
Notice that the solution \eqref{Q1*} and, hence, also the limit of $AdS_5$ Chern--Simons gravity, do not admit the asymptotic flatness for large $r$. 

\subsection{Generalization to $d+1$ Dimensions}
\label{GEND}

We now extend the previous results to $d+1$ dimensions, finding out analytical solutions under the assumption $P(r) = 1/Q(r)$. In this way, not all the solutions provided in Sec.~\ref{CSSS} can be recovered under given limits. As a starting point, the 5-dimensional line element~\eqref{INT5} can be extended to:
\begin{equation}
\label{sphsymmet}
ds^2 = P(r)^2 dt^2 - Q(r)^2 dr^2 - r^2 d\Omega_{d-1}^2\,, 
\end{equation}
By means of this choice, the Ricci scalar and the Gauss--Bonnet term can be recast in terms of $P$ and $Q$, as:
\begin{eqnarray}
\label{GBterm_spherical}
\G^{(d+1)} = \frac{(d-2)(d-1)}{r^4 P Q^5} && \left\{(d-3) P \left(Q^2-1\right) \left[(d-4)
   Q^3-(d-4) Q+4 r Q'\right]  \right.  \nonumber
   \\
   &&\left.  -4 r \left[ (d-3) Q^3 P'+r
  Q^3 P'' - (d-3)Q P'- r Q P'' -r Q^2 P' Q'+3 r P'
   Q'\right]  \right\}\,,
 \end{eqnarray}
 \begin{equation}
  \R^{(d+1)} = \frac{2 r \left\{Q \left[(d-1) P'+r P''\right]-r P' Q'\right\}+(1-d) P \left[(d-2) Q^3+(2-d) Q+2 r Q'\right]}{r^2 P Q^3}\,.
 \end{equation}
The Lovelock point-like Lagrangian only containing $\R, \G$ and the cosmological constant $\alpha_0$, can be obtained by integrating out the second derivatives. After some basic computations, it takes the form
\begin{eqnarray}
\Lagr^{(d+1)} = \frac{r^{d-5} P \prod_{j=2}^{d-1} (\sin \theta_{j-1})^{d-j}}{Q^4} && \left\{\alpha_0 r^4 Q^5 -  \alpha_1 (d-1) r^2 Q^2 [(d-2) Q(Q^2-1) + 2 r Q']  \right.  \nonumber
\\
&& \left. + \alpha_2 (d-3)(d-2)(d-1)(Q^2-1) [(d-4) Q(Q^2-1) + 4 r Q'] \right\}\,,
\label{Lovelockd+1Lagr}
\end{eqnarray}
whose equations of motion with respect to $P(r)$ and $Q(r)$ are, respectively:
\begin{eqnarray}
&& \alpha_2 (d-3) (d-2) (d-1) \left(Q^2-1\right) \left[(d-4) Q \left(Q^2-1\right)+4 r Q'\right)+\alpha_0 r^4 Q^5 \nonumber
\\
&& -\alpha_1 (d-1) r^2 Q^2 \left[(d-2) Q \left(Q^2-1\right)+2 r Q'\right] = 0\,, \nonumber
\\ \nonumber \\
&& 2 (d-1) r P' \left\{Q^2 \left[2 \alpha_2 (d-3)(d-2)-\alpha_1 r^2\right]-2 \alpha_2 (d-3)(d-2)\right\} \nonumber \\
&& -P \left\{(d-1)(d-2) Q^2 \left[\alpha_1 r^2-2 \alpha_2 (d-3)(d-4)\right] +\alpha_2 \left(d^4-10 d^3+35 d^2-50
   d+24\right) \right. \nonumber
\\
&&\left. +Q^4 \left[-\alpha_1 (d-1)(d-2) r^2+ \alpha_2 \left(d^4-10 d^3+35 d^2-50 d+24\right)+\alpha_0 r^4\right]\right\} = 0\,.
\end{eqnarray}
The general solution of the Euler--Lagrange equations can be found analytically by imposing $P(r) = Q(r)^{-1}$. Under this assumption, the spherically symmetric solution for $P(r)$ and $Q(r)$ reads as
\begin{eqnarray}
P(r)^2 = 1 \pm \frac{1}{r^{d/2 -2}} \sqrt{\frac{c_1}{6 \alpha_2 \binom{d-1}{d-4}}+r^d
   \left(\frac{\alpha_1^2}{16 \alpha_2^2\binom{d-2}{d-4}^2}- \frac{\alpha_0}{24 \alpha_2 \binom{d}{d-4}}\right)}-\frac{\alpha_1}{4 \alpha_2 \binom{d-2}{d-4}} \, r^2\,,  \nonumber
\end{eqnarray}
\begin{eqnarray}
Q(r) = \frac{1}{P(r)}\,.
\label{Q1}
\end{eqnarray}
When Lovelock's parameters $\alpha_i$ are chosen as in Refs. \cite{Crisostomo:2000bb, Banados:1993ur}, the term in the brackets vanishes and the solution \eqref{Q1} can be recast as:
\begin{equation}
P(r)^2 = 1 \pm \frac{1}{l (d-2)(d-3) r^{d/2 -2}} \sqrt{\frac{c_1}{(d-1)}}+\frac{r^2}{l^2}\,,  \nonumber
\end{equation} 
which reduces to the static spherically symmetric solution provided in Ref. \cite{Crisostomo:2000bb} by an appropriate rescaling of the integration parameter $c_1$. On the other hand, with the same choice for Lovelock's parameters, solution in Ref. \cite{Banados:1993ur} is recovered in six dimensions, where Eq. \eqref{Q1} becomes:
\begin{equation}
P(r)^2 = 1 \pm \frac{1}{12l} \sqrt{\frac{c_1}{r}}+\frac{r^2}{l^2}\,. \nonumber
\end{equation}
When $\alpha_0 = 0$, Eq. \eqref{Q1} reduces to
\begin{equation}
P(r)^2 = 1 \pm \frac{1}{r^{d/2 -2}} \sqrt{\frac{c_1}{6 \alpha_2 \binom{d-1}{d-4}}+
 \frac{\alpha_1^2}{16 \alpha_2^2\binom{d-2}{d-4}^2} \, r^d}-\frac{\alpha_1}{4 \alpha_2 \binom{d-2}{d-4}} \, r^2\,,
\label{Q2}
\end{equation}
which is nothing but the solution found in Refs. \cite{Crisostomo:2000bb, Wiltshire:1988uq, Whitt:1988ax} with a proper redefinition of the integration constants.
Both in Eq. \eqref{Q1} and in Eq. \eqref{Q2}, the horizon cannot be found analytically for any value of the integration constant $c_1$. However, it can be found after expanding the metric up to the first order. By means of the definitions
\begin{equation}
\tilde{c}_1 \equiv \frac{c_1}{6 \alpha_2 \binom{d-1}{d-4}} \quad \alpha \equiv \left(\frac{\alpha_1^2}{16 \alpha_2^2\binom{d-2}{d-4}^2}- \frac{\alpha_0}{24 \alpha_2 \binom{d}{d-4}}\right) \quad \Lambda \equiv -\frac{\alpha_1}{4 \alpha_2 \binom{d-2}{d-4}},
\end{equation}
Eq.~\eqref{Q1} can be rewritten as
\begin{equation}
    P(r)^2 = 1 \pm \frac{1}{r^{d/2 -2}} \sqrt{\tilde{c}_1 + \alpha r^d} +\Lambda \, r^2\,,
\end{equation}
so that under the assumption $\alpha r^d \ll \tilde{c}_1$, $P(r)^2$ becomes
\begin{equation}
P(r)^2 = 1 - r^{2-\frac{d}{2}} \sqrt{c_1} \left(1 + \frac{\alpha}{2 \tilde{c}_1} r^d \right) + \Lambda r^2 = \Lambda\left[ \frac{\frac{r^{\frac{d}{2}-2}}{\Lambda} - \frac{\sqrt{c_1}}{\Lambda}\left(1 + \frac{\alpha}{2 \tilde{c}_1} r^d \right) + r^{\frac{d}{2}} }{r^{\frac{d}{2}-2}} \right].
\end{equation}
When $\Lambda \gg r^{\frac{d}{2}-2}$, the horizon can be computed analytically, providing
\begin{equation}
r_H = \left[\frac{2}{\alpha} (-\tilde{c}_1) \left(1 + d^{d/2} \right)\right]^{1/d} = \left[\frac{8 \alpha_2 d(d-1)(d-2)^2(d-3)^2}{ \alpha_1^2 d(d-1)- 4 \alpha_0 \alpha_2 (d-2)(d-3)}(-c_1) \left(1 + d^{d/2} \right)\right]^{1/d}\,.
\end{equation}
Considering that the ratio $\alpha_2/\alpha_1^2$ must be dimensionless, the constant $c_1$ must have a mass dimension and the horizon is proportional to 
\begin{equation}
    r_H \sim M^{1/d}\,.
\end{equation}
The Bekenstein--Hawking entropy $\mathcal{S}$, therefore, can be written in terms of $r_H$ as:
\begin{equation}
    \mathcal{S}  =\frac{\pi^{\frac{d}{2}} }{2 \Gamma(\frac{d}{2})} \left[\frac{8 \alpha_2 d(d-1)(d-2)^2(d-3)^2}{ \alpha_1^2 d(d-1)- 4 \alpha_0 \alpha_2 (d-2)(d-3)} \left(1 + d^{d/2} \right)\right]^{\frac{d-1}{d}}\,  M^{\frac{d-1}{d}} ,
    \label{BHLOV}
\end{equation}
where $\Gamma$ is the Euler Gamma function. Comparing Eq.~\eqref{BHLOV} with the entropy of a CFT in $d+1$ dimensions, namely~\cite{Myung:2001ab, Aharony:1998tt}
\begin{equation}
\mathcal{S}_{CFT} \sim M^{\frac{d}{d+1}},
\end{equation}
we see that a $D$ dimensional CFT behaves like a $D+1$ dimensional AdS-invariant theory, in terms of entropy scaling. This is directly linked to the AdS/CFT correspondence and does not hold when $\alpha_0 = 0$. 
Following the same prescription, now we find exact solutions when $\alpha_2 = 0, \, \alpha_1 = 0$ and show that the entropy scales as that of a CFT only when $\alpha_0 \neq 0$.

The $\alpha_2 = 0$ limit must be found separately and provides the well known high-dimensional Schwarzschild--de Sitter solution
\begin{equation}
P(r)^2 = \frac{1}{Q(r)^2} = 1+ \frac{c_1}{r^{d-2}}-\frac{\alpha_0}{\alpha_1 d (d-1)} r^2\,,
\label{GRd+1}
\end{equation}
with a proper redefinition of the constant $c_1$. 
The horizon can be found under the assumption $\alpha_0 r^2 \gg 1$, where the element $P(r)^2$ can be approximated to
\begin{equation}
P(r)^2 = 1+ \frac{c_1}{r^{d-2}}-\frac{\alpha_0}{\alpha_1 d (d-1)} r^2\, = \frac{\alpha_0}{\alpha_1 d (d-1)} \left(\frac{\frac{\alpha_1 d (d-1)}{\alpha_0} \frac{r^{d}}{r^2} + \frac{\alpha_1 d (d-1)}{\alpha_0} \, c_1 - r^d}{r^{d-2}} \right) \sim \frac{\alpha_0}{\alpha_1 d (d-1)} \left( \frac{\frac{\alpha_1 d (d-1)}{\alpha_0} \, c_1 - r^d}{r^{d-2}}\right),
\end{equation}
so that the horizon is
\begin{equation}
r_H \sim \left( \frac{c_1 \alpha_1 d (d-1)}{\alpha_0}\right)^{1/d}\,.
\end{equation}
Setting $\alpha_0 = 0$, the $d+1$ dimensional generalization of Schwarzschild radius can be computed without approximations and turns out to be
\begin{equation}
r_H = (- c_1)^{\frac{1}{d-2}}\,.
\end{equation}
Identifying the constant $c_1$ with $-M$, as previously discussed, we notice that the Bekenstein--Hawking entropy $\mathcal{S}$ scales as 
\begin{equation}
    \mathcal{S} \sim \frac{ \pi^{\frac{d}{2}} }{2 \Gamma(\frac{d}{2})} \left( \frac{\alpha_1 d (d-1)}{\alpha_0}\right)^{\frac{d-1}{d}} \, M^{\frac{d-1}{d}},
       \label{BHADS}
\end{equation}
for $\alpha \neq 0$ and as
\begin{equation}
     \mathcal{S} = \frac{\pi^{\frac{d}{2}} }{2 \Gamma(\frac{d}{2})} M^{\frac{d-1}{d-2}},
     \label{BHGR}
\end{equation}
for $\alpha_0 = 0$. This means that the AdS/CFT correspondence holds as long as $\alpha_0 \neq 0$, as expected. 

Finally, assuming $\alpha_1 = 0$, the Lagrangian reduces to a sum of the Gauss--Bonnet term and the cosmological constant, providing~\cite{Bajardi:2019zzs}:
 \begin{equation}
P(r)^2 = 1 \pm \frac{1}{r^{d/2 -2}} \sqrt{\frac{c_1}{6 \alpha_2 \binom{d-1}{d-4}}-r^d \frac{\alpha_0}{24 \alpha_2 \binom{d}{d-4}}}\,.  \nonumber
\label{GBsph}
\end{equation}  
It is worth stressing out that the above solution turns out to be trivial in less than 5 dimensions, as expected from the topological nature of $\G$. Moreover, it only holds for $\alpha_0 \neq 0$. Under the same approximations as Lovelock case, the first-order expansion of $P(r)^2$ yields the horizon:
\begin{equation}
r_H \sim \left(\frac{2}{\alpha_0}\right)^{1/d} (-c_1)^{1/d}\,,
\end{equation}
which means that the Bekenstein--Hawking entropy is proportional to
\begin{equation}
\mathcal{S} \sim \frac{\pi^{\frac{d}{2}} }{2\Gamma(\frac{d}{2})} \left(\frac{2}{\alpha_0}\right)^{\frac{d-1}{d}}  M^{\frac{d-1}{d}}.
\end{equation}
Also here, the AdS/CFT correspondence is recovered, unlike the case of Gauss--Bonnet gravity with $\alpha_0 = 0$. As a matter of fact, when assuming $\alpha_0 = 0$ from the beginning, the equations of motion yield
\begin{equation}
P(r)^2 = \frac{1}{Q(r)^2} = \left(1 + \frac{c_{1}}{r^{\frac{d}{2}-2}}\right),
\end{equation}
admitting an horizon at 
\begin{equation}
r_H = (-c_1)^{\frac{2}{d-4}}\,,
\end{equation}
whose corresponding entropy scales as
\begin{equation}
  \mathcal{S} = \frac{\pi^{\frac{d}{2}} }{2 \Gamma(\frac{d}{2})} M^{\frac{2d -2}{d-4}}.
\end{equation}
Gauss--Bonnet gravity is deeply studied in spherically symmetric and cosmological backgrounds. For instance, higher-dimensional NUT black hole solutions are discussed in Refs.~\cite{Mukherjee:2020lld, Mukherjee:2021erg}, where the authors find out solutions of pure Gauss--Bonnet $\Lambda$-vacuum equation describing a black hole with NUT charge.

As a final remark, notice that Eq.~\eqref{Q1}, together with the corresponding subcases above discussed, is the generalization of Eq.~\eqref{Q1*} to $d+1$ dimensions. 

\section{Discussion and Conclusions}
\label{Concl}

Cosmological and spherically symmetric solutions for Lovelock and $AdS_5$ Chern--Simons gravity have been found, also discussing the extension of the 5-dimensional Lovelock Lagrangian to $D$ dimensions.
In particular, we have demonstrated that $AdS_5$ Chern--Simons theory is recovered as a particular case of Lovelock gravity. Specifically, the Chern--Simons gravity is a topological theory capable of fixing some issues which plague GR in view of  Quantum Gravity. In particular, Chern--Simons approach  well fits the formalism of other fundamental interactions, so it can be derived as an effective  theory from Supergravity or  Strings. 

Here, we showed that the 5-dimensional Lovelock gravity admits exact cosmological solutions, with exponential scale factors. Moreover, after introducing the fifth dimension $x_4$ labeled by the function $b(t)$, it turns out that under the assumption $a(t) = b(t)$ the field equations yield other exact solutions. Depending on the values of the coupling constants, Lovelock cosmology admits both de Sitter-like solutions and bouncing space-times. The restriction to $AdS_5$ Chern--Simons gravity sets the values of the coupling constants such that only exponential scale factors with real exponents are allowed. When generalizing the treatment to $D$ dimensions, we also consider the spatial curvature $k$ in the  line element. We found the solutions of the corresponding equations of motion and showed that, in the limit $k = 0$, only exponential solutions occur. In Table~\hyperref[TableI]{I} the cosmological solutions are summarized.

On the other hand, the study of Lovelock gravity in a spherically symmetric background, shows that several solutions can be obtained from the general $D$-dimensional Lagrangian~\eqref{Lovelockd+1Lagr}. In 5 dimensions, analytical solutions can be found without imposing any relation between the components of the line elements $P(r)$ and $Q(r)$. Similarly to the cosmological case, AdS-invariant Chern--Simons gravity can be obtained through an appropriate choice of the coupling constants. We extend the treatment to $D$ dimensions, and after imposing $P(r) = 1/Q(r)$, we find exact solutions. It is worth stressing that not all the solutions are physically relevant and, often, a careful selection of the free parameters is necessary. On the other hand, in some other cases, physical solutions cannot be recovered regardless of the values assumed by the integration constants. The spherically symmetric solutions are summarized in Table~\hyperref[TableII]{II}. 

As we can infer from Table~\hyperref[TableII]{II}, only the 5-dimensional case of pure Gauss--Bonnet gravity leads to mathematically trivial solutions. The main purpose of this work is to show that most of the theories coming from the general Lovelock action can be extended in more than 4 dimensions. This is worth \emph{e.g.} in view of the AdS/CFT correspondence, according to which the $D$-dimensional AdS-invariant Lovelock action should be equal to a $(D-1)$-dimensional CFT. Although in 4 dimensions GR perfectly fits the current observations at Solar System scales, in more dimensions there are several other candidate theories capable of describing  the observed cosmological phenomenology as effects of dimensional reduction and higher-order dynamics. In general, the dark side issue could be related to these processes.  In particular, cosmological and spherically symmetric solutions of pure Gauss--Bonnet gravity hold only for $D \ge 5$, unless the coupling constant $\alpha_2$ diverges when $D = 4$. This idea was recently proposed \emph{e.g.} in Refs.~\cite{Gurses:2020rxb, Glavan:2019inb}, where the authors deal with a 4-dimensional theory in which the Gauss--Bonnet term contributes to the dynamics. In future works we aim to find more general solutions, even in presence of matter. 

\newpage
\begin{center} Table I: \, \emph{Cosmological Solutions in Lovelock and $AdS_5$ Chern--Simons Gravity}
\label{TableI}
\end{center}

\resizebox{17.5cm}{!}{
\begin{centering}
\begin{tabular}{|c |c c c |c |c |c|}\hline\hline
\multicolumn{1}{|c|}{\textbf{Case}} & \textbf{$\alpha_0$} & \textbf{$\alpha_1$}
& \textbf{$\alpha_2$} & $k$ & \textbf{$\text{Scale \, Factor}$} & Dimension \\ \hline 
&  & &  & &  &\\
Einstein--de Sitter& $\neq 0$ & $\neq 0$ & $0$  & $\neq 0$  & $\displaystyle a(t) = \pm \sqrt{\frac{\alpha_1 k \, d (d-1)}{\alpha_0 - \alpha_0 \coth ^2\left[\sqrt{\alpha_0} \left(c_1+\frac{t}{\sqrt{\alpha_1 (d-1) d}}\right)\right]}} $  & Any \\ 
& & & & & & \\
\cline{5-6}
& & & & & & \\
& & & & 0 & $\displaystyle a(t) = a_0  e^{\pm \sqrt{\frac{\alpha_0}{\alpha_1 d(d-1)}}\, t} $ & \\
&  & &  & & & \\
\cline{6-7}
&  & &  & & & \\
&  & &  & & $\displaystyle a(t) = a_0  e^{\pm \sqrt{\frac{\alpha_0}{12\alpha_1}}\, t}$ & 5 \\ 
& & & & & & \\
\cline{1-7}
&  & &  & & & \\
Pure Gauss--Bonnet \,\, & $\neq 0$ & $0$ & $ \neq 0 $ & 0 & $\displaystyle a(t) = a_0 \exp\left\{\pm \sqrt[4]{\frac{-\alpha_0}{d(d-1)(d-2)(d-3)  \alpha_2 }} \, t\right\} \,\,\, a(t) = b(t)$ & Any \\ 
&  & &  & & & \\
\cline{6-7}
&  & &  & & & \\
&  & &  &  & $\displaystyle a(t) = a_0 \exp\left\{\pm \sqrt[4]{\frac{-\alpha_0}{24  \alpha_2 }} \, t\right\}$ & 5 \\
&  & &  & &  &\\
\cline{2-7}
&  & &  & & & \\
& $0$ & $0$ & $ \neq 0 $ & $\neq 0$& $a(t) = \sqrt{-k} \, t$  & Any \\ 
&  & &  & & & \\
\cline{5-6}
&  & &  & & & \\
& & & & 0 & $\displaystyle a(t) \sim \text{Const.} $ & \\
&  & &  & & & \\
\cline{6-7}
&  & &  & & & \\
&  & &  & & $\displaystyle a(t) \sim \text{Const.} $ & 5 \\
&  & &  & & & \\
\cline{1-7}
&  & &  & & & \\
Lovelock & $\neq 0$ & $\neq 0$ & $\neq 0$ & 0 & $\displaystyle \qquad \qquad a(t)= a_0 \exp\left\{\pm \sqrt{\frac{2 \alpha_0}{\pm \sqrt{(d-1) d \left[\alpha_1^2 (d-1) d-4 \alpha_0 \alpha_2 (d-3) (d-2)\right]}+\alpha_1 d(d-1)}} \, t\right\} $ & Any \\
&  & &  & & & \\
\cline{6-7}
&  & &  & & & \\
&  & & & & $\displaystyle a(t) = a_0 \exp\left\{\pm \sqrt{\frac{\alpha_0}{\pm 2 \sqrt{9 \alpha_1^2-6 \alpha_0 \alpha_2}+6 \alpha_1}} \,  t \right\} $ & 5 \\ 
&  & &  & & & \\
\cline{2-7}
&  & &  & & & \\
 & $0$ & $\neq 0$ & $\neq 0 $ & $\neq 0$ & $\displaystyle a(t) = \pm \sqrt{\frac{- \alpha_2 k (d-3) (d-2)}{\alpha_1}} \sinh \left[\sqrt{\alpha_1} \left(\frac{t}{\sqrt{\alpha_2 (d-3)(d-2)}}+c_1\right)\right]$  & Any \\ 
&  & &  & & & \\
\cline{5-6}
&  & &  & & & \\
& & & & 0 & $\displaystyle a(t) = a_0 \exp\left\{\pm \sqrt{\frac{\alpha_1}{\alpha_2 \left(d-2\right)(d-3)}}\, t\right\}$ & \\
&  & &  & & & \\
\cline{6-7}
&  & &  & & & \\
&  & &  & & $\displaystyle a(t) = a_0 \exp\left\{\pm \sqrt{\frac{\alpha_1}{2 \alpha_2}}\, t\right\}$ & 5 \\ 
&  & &  & & & \\
\cline{1-6}
&  & &  & & & \\
Chern--Simons &$\displaystyle \frac{1}{5l^4}$ & $\displaystyle \frac{2}{3l^2}$ & $\displaystyle 1$ & 0 & $ a(t) = a_0 \exp\left\{\pm \frac{1}{l} \sqrt{\frac{1}{6}\left(1\pm \sqrt{ \frac{7}{10}}\right)} \,\, t\right\}  $&  \\
&  & &  & & & \\
   
\hline
\end{tabular}
\end{centering}
}

\newpage
 \begin{center} 
Table II: \, \emph{Spherically Symmetric Solutions in Lovelock and $AdS_5$ Chern--Simons Gravity}
\label{TableII}
\end{center}
\resizebox{17.5cm}{!}{
\begin{centering}
\begin{tabular}{ |c |c c c |c |c|} \hline  
 \multicolumn{1}{|c|}{\textbf{Case}} & \textbf{$\alpha_0$} & \textbf{$\alpha_1$}
& \textbf{$\alpha_2$} & \textbf{$P(r)^2$, \,\, $Q^2(r)$} & Dimension \\  \hline 
&  & &  & & \\
Einstein--de Sitter& $\neq 0$ & $\neq 0$ & $0$ & $\displaystyle P(r)^2 = 1/Q(r)^2 = 1+\frac{c_1}{r^{d-2}}-\frac{\alpha_0}{\alpha_1d (d-1)} r^2$  & Any \\ 
&  & &  & & \\ 
\cline{5-6}
&  & &  & & \\
&  & &  & $\displaystyle P(r)^2 = 1/Q(r)^2 = 1 + \frac{c_1}{r^2} - \frac{\alpha_0}{12 \alpha_1} r^2$  & 5 \\ 
&  & &  & & \\ \cline{1-6}
&  & &  & & \\
Pure Gauss--Bonnet \,\, & $\neq 0$ & $0$ & $ \neq 0 $ & $\displaystyle P(r)^2 = 1/Q(r)^2  = 1 \pm \frac{1}{r^{d/2 -2}} \sqrt{\frac{c_1}{6 \alpha_2 \binom{d-1}{d-4}}-r^d \frac{\alpha_0}{24 \alpha_2 \binom{d}{d-4}}} $ & Any  \\  
&  & &  & & \\ 
\cline{5-6}
&  & &  & & \\
&  & &  &  $ \qquad \qquad \displaystyle P(r)^2 = 1/Q(r)^2 = 1\pm \sqrt{1+c_1 - \frac{\alpha_0}{24 \alpha_2} r^4}$ & 5 \\
&  & &  & & \\
\cline{5-5}
&  & &  & & \\
& & & & $\qquad \qquad  \displaystyle P^2(r) = P_0^2 \sqrt{48 \alpha_2 \left(2 c_2+1\right) \mp 4 \sqrt{6} \sqrt{\alpha_2 \left(24 \alpha_2 +96 \alpha_2  c_2-\alpha_0 r^4\right)}-\alpha_0 r^4} \qquad \displaystyle Q^2(r) = \frac{2 \left(12 \alpha_2 \pm \sqrt{6} \sqrt{\alpha_2 \left(24 \alpha_2+96 \alpha_2 c_2-\alpha_0
   r^4\right)}\right)}{\alpha_0 r^4-96 \alpha_2 c_2}$& \\ 
   &  & &  & & \\ \cline{2-6}
   &  & &  & & \\
 & $0$ & $0$ & $ \neq 0 $ & $\displaystyle P(r)^2 = 1/Q(r)^2 = 1 + \frac{c_{1}}{r^{\frac{d}{2}-2}}$ & Any \\ 
&  & &  & & \\
\cline{5-6}
&  & &  & & \\
&  & &  & \text{Const.} & 5 \\
&  & &  & & \\
\cline{1-6}
&  & &  & & \\
Lovelock & $\neq 0$ & $\neq 0$ & $\neq 0$ & $\quad \displaystyle P(r)^2 = 1/Q(r)^2 = 1 \pm \frac{1}{r^{d/2 -2}} \sqrt{\frac{c_1}{6 \alpha_2 \binom{d-1}{d-4}}+r^d
   \left(\frac{\alpha_1^2}{16 \alpha_2^2\binom{d-2}{d-4}^2}- \frac{\alpha_0}{24 \alpha_2 \binom{d}{d-4}}\right)}-\frac{\alpha_1}{4 \alpha_2 \binom{d-2}{d-4}} \, r^2$ & Any \\ 
   &  & &  & & \\
   \cline{5-6}
    &  & &  & & \\
&  & &  & $\qquad \qquad P(r)^2 = 1/Q(r)^2 = \displaystyle 1-\frac{\alpha_1 r^2}{4
   \alpha_2} \pm \frac{\sqrt{3 r^{4} \left(3 \alpha_1^2-2 \alpha_0
   \alpha_2\right)+6 \alpha_2 c_1 }}{12 \alpha_2}$ & 5 \\ 
   &  & &  & & \\
   \cline{5-5}
    &  & &  & & \\
   & & & & $ \displaystyle P(r)^2=P_0^2 \sqrt{-4 c_1+\alpha_0 r^4-12 \alpha_1 r^2}
   \left[\frac{3 \sqrt{r^4 u w+2 u x}+\sqrt{3} \left(\alpha_0 x-r^2 w (-3
   \alpha_1+z)\right)}{-6 \alpha_1+\alpha_0 r^2+2
   z}\right]^{y/2} \left[\frac{6 \alpha_1-\alpha_0 r^2+2 z}{3
   \sqrt{r^4 v w+2 v x}+\sqrt{3} \left(r^2 w (z+3 \alpha_1)+\alpha_0
   x\right)}\right]^{s/2}$ & \\ 
    &  & &  & & \\
   & & & & $\displaystyle Q(r)^2 = \frac{-3 \alpha_1 r^2+12 \alpha_2 \pm \sqrt{3} \sqrt{8 c_1 \alpha_2 -2 \alpha_0 \alpha_2 r^4+3 \alpha_1^2 r^4+48 \alpha_2^2}}{\frac{\alpha_0}{2} r^4-6 \alpha_1 r^2 -2 c_1}  $ & \\ 
   &  & &  & & \\
   \cline{2-6}
    &  & &  & & \\
 & $0$ & $\neq 0$ & $\neq 0$ & $\quad \displaystyle P(r)^2 = 1/Q(r)^2 = 1 \pm \frac{1}{r^{d/2 -2}} \sqrt{\frac{c_1}{6 \alpha_2 \binom{d-1}{d-4}}+\frac{\alpha_1^2}{16 \alpha_2^2\binom{d-2}{d-4}^2} \, r^d}-\frac{\alpha_1}{4 \alpha_2 \binom{d-2}{d-4}} \, r^2$ & Any \\ 
&  & &  & & \\
\cline{5-6}
    &  & &  & & \\
&  & & & $\qquad \qquad \displaystyle P(r)^2 = 1/Q(r)^2 =  1-\frac{\alpha_1 r^2}{4
   \alpha_2} \pm \frac{\sqrt{9\alpha_1^2 r^{4}+6 \alpha_2 c_1 }}{12 \alpha_2}$ & 5 \\ 
   &  & &  & & \\
   \cline{5-5}
    &  & &  & & \\
   & & & & $ \displaystyle P(r)^2 = P_0^2 \left(2 c_1+ \alpha_1 r^2\right) \sqrt{\frac{\sqrt{16 \alpha_2 (\alpha_2+c_1)+\alpha_1^2 r^4}+\alpha_1 r^2}{8 \alpha_2^2+2 \alpha_2 \left(\sqrt{16 \alpha_2 (\alpha_2+c_1)+\alpha_1^2 r^4}+4 c_1\right)+c_1 \left(\sqrt{16 \alpha_2
   (\alpha_2+c_1)+\alpha_1^2 r^4}-\alpha_1 r^2\right)}}\,,$ & \\ 
    &  & &  & & \\ 
     & & & & $\displaystyle Q(r)^2 = \frac{-4 \alpha_2 - \sqrt{16 \alpha_2 (\alpha_2 +c_1)+\alpha_1^2 r^4}+\alpha_1 r^2}{4 c_1+2 \alpha_1 r^2} $ & \\
   &  & &  & & \\
   \cline{5-5}
     &  & &  & & \\
     & & & & $ \displaystyle P(r)^2 = P_0^2 \sqrt{\frac{8 \alpha_2^2+2 \alpha_2 \left(\sqrt{16 \alpha_2 (\alpha_2 + c_1)+\alpha_1^2 r^4}+4 c_1\right)+c_1 \left(\sqrt{16
   \alpha_2 (\alpha_2+c_1)+\alpha_1^2 r^4}-\alpha_1 r^2\right)}{\sqrt{16 \alpha_2 (\alpha_2+c_1)+\alpha_1^2 r^4}+\alpha_1 r^2}}$ & \\
   &  & &  & & \\ 
    & & & & $\displaystyle Q(r)^2 = \frac{-4 \alpha_2 + \sqrt{16 \alpha_2 (\alpha_2 +c_1)+\alpha_1^2 r^4}+\alpha_1 r^2}{4 c_1+2 \alpha_1 r^2} $ & \\
   &  & &  & & \\
   \cline{1-5}
    &  & &  & & \\
Chern--Simons &$\displaystyle \frac{1}{5l^4}$ & $\displaystyle \frac{2}{3l^2}$ & $\displaystyle 1$ & $ \qquad \qquad \displaystyle P(r)^2 = 1/Q(r)^2 =1-\frac{r^2}{6 l^2}\pm  \sqrt{\frac{7 r^{4}}{360 l^4}+\frac{ c_1}{24}}$&  \\ 
&  & &  & & \\
\cline{5-5}
    &  & &  & & \\
& & & & $\displaystyle P(r)^2=P_0^2 \sqrt{-4 c_1 +\frac{r^4}{5 l^4}-\frac{8 r^2}{l^2}}
   \left[\frac{\sqrt{3} \left(\frac{x}{5 l^4}-r^2 w
   \left(-\frac{2}{l^2}+z\right)\right)+3 \sqrt{u \left(r^4 w+2 x\right)}}{\frac{r^2}{5
   l^4}-\frac{4}{l^2}+2 z}\right]^{y/2} 
 \left[\frac{-\frac{r^2}{5 l^4}+\frac{4}{l^2}+2
   z}{\sqrt{3} \left(\frac{x}{5 l^4}+r^2 w \left(z+\frac{2}{l^2}\right)\right)+3 \sqrt{v
   \left(r^4 w+2 x\right)}}\right]^{s/2}$ & \\ 
   &  & &  & & \\
& & & & $\displaystyle Q(r)^2 = \frac{\sqrt{24 c_1 + \frac{14 r^4}{5 l^4}+144}-\frac{2
   r^2}{l^2}+12}{-2 c_1 +\frac{r^4}{10 l^4}-\frac{4 r^2}{l^2}} $ & \\  
   &  & &  & & \\
   \hline

\end{tabular}
\end{centering}
}

\section*{Acknowledgments}

The Authors acknowledge the support of {\it Istituto Nazionale di Fisica Nucleare} (INFN) ({\it iniziative specifiche} GINGER, MOONLIGHT2, QGSKY, and TEONGRAV). This paper is based upon work from COST action CA15117 (CANTATA), COST Action CA16104 (GWverse), and COST action CA18108 (QG-MM), supported by COST (European Cooperation in Science and Technology).

\appendix

\begin{appendix}
\section{The Cartan Structure Equations}
\label{APPA}

This Appendix aims to recall some basic concepts of differential forms and Cartan's structure equations used in Secs.~\ref{LOVCOS} and~\ref{CSSS}. Let $\mathcal{P}$ and $\mathcal{Q}$ be a $n$-form and a $m$-form, respectively; the ``exterior product'' $T$ is the $(m+n)$-form given by 
\begin{equation}
T= \mathcal{P} \land \mathcal{Q} = \mathcal{P}_{[\alpha_1,\alpha_2,...\alpha_n} \mathcal{Q}_{\alpha_{n+1},\alpha_{n+2},...,\alpha_{n+m}]} \; dx^{\alpha_1}\land dx^{\alpha_2} \land.. \land dx^{\alpha_{m+n}}\,.
\end{equation}
The ``exterior derivative'' accounts for the exterior product (or wedge product) between the 1-form derivative $d$ and the $n$-form $\mathcal{P}$:
\begin{equation}
d\mathcal{P}= \partial_{[\alpha_1} P_{\alpha_2 \alpha_3...\alpha_n]} \; dx^{\alpha_1} \land dx^{\alpha_2}\land ...\land dx^{\alpha_n}\,.
\label{external derivative}
\end{equation}
By defining the Hodge dual $\star \mathcal{P}$ of a $n$-form $\mathcal{P}$ as the $m-n$ form
\begin{equation}
\star \mathcal{P} = \frac{1}{(m-n)!}\mathcal{P}^{\alpha_1,\alpha_2,...\alpha_n} \sqrt{|g|}\epsilon_{\alpha_1,\alpha_2,...\alpha_m}\; dx^{\alpha_{n+1}} \land dx^{\alpha_{n+2}} \land ... \land dx^{\alpha_{m}}\,,
\label{duale di Hodge}
\end{equation} 
it automatically follows that
\begin{equation}
\star 1 = (-1)^{D-1} \sqrt{|g|} d^D x\,.
\end{equation}
Therefore, the wedge product can be written as
\begin{equation}
dx^{\alpha_1} \land dx^{\alpha_2} \land...\land dx^{\alpha_D} = \epsilon^{\alpha_1, \alpha_2 ...\alpha_D} d^D x\,.
\label{relazione prodotto esterno}
\end{equation}
The differential forms can be also expressed in coordinates representations by means of the vielbein basis (tetrad in 4 dimensions). Starting from the definition of vielbein field, \emph{i.e.},
\begin{equation}
g_{\mu \nu} = \partial_\mu x^a \partial_\nu x^b  \eta_{ab} \equiv e^a_\mu e^b_\nu \eta_{ab} \;,
\end{equation}
it is possible to define $e^a = e^a_\mu dx^\mu$, so that a generic $n$-form $\mathcal{P}$ can be written as:
\begin{equation}
\mathcal{P} = \mathcal{P}_{[\alpha_1, \alpha_2...\alpha_n]} dx^{\alpha_1} \land dx^{\alpha_2} \land ... \land dx^{\alpha_n} = \mathcal{P}_{[a_1, a_2...a_n]} e^{a_1} \land e^{a_2} \land ... \land e^{a_n}\,.
\end{equation}
Considering a Lorentz transformation of the form $\Lambda = \displaystyle e^{\frac{i}{2} \omega^{ab} J_{ab}}$, with $\omega^{ab}$ being the 1-form Lorentz connection, the covariant derivative of the $n$-form $\mathcal{P}$ is the $n+1$-form 
\begin{equation}
D \mathcal{P} = d\mathcal{P} - \frac{i}{2} \omega^{ab} J_{ab} \mathcal{P}\,.
\label{DP}
\end{equation}
The covariant derivative of Eq.~\eqref{DP} provides
\begin{equation}
D \land D \mathcal{P} = - \frac{i}{2} R^{ab} J_{ab} \mathcal{P} \;\;\;\;\;\;\;\; R^{ab}= d \omega^{ab} + \omega^a_c \land \omega^{cb}\,,
\label{structural eq1}
\end{equation}
where $R^{ab}$ is the so called curvature 2-form. Considering the antisymmetric part of vielbein postulation, namely
\begin{equation}
D_{[\mu} e^a_{\nu]} = \partial_{[\mu} e^a_{\nu]} + \omega^a_{[\mu \nu]}= Q^a_{\mu \nu}\,,
\label{antitetrad}
\end{equation}
the so called ``contorsion tensor'' $Q^a_{\mu \nu}$ arises. By means of Eq.~\eqref{antitetrad}, the torsion 2-form can be defined as:
\begin{equation}
T^a = De^a = de^a + \omega^a_b \land e^b\,.
\label{structural eq2}
\end{equation}
Eqs.~\eqref{structural eq1} and~\eqref{structural eq2} are the so-called Cartan structure equations. Notice that while the curvature 2-form is the gauge field associated to Lorentz transformation, torsion is the gauge field associated to local translations. Setting to zero the Lorentz connections, gravity is described by the torsion only, and the corresponding theory is the Teleparallel Gravity, mentioned in the Introduction.

\section{Values of Constants of Sec.~\ref{CSSS}}
\label{APPB}

\begin{center}
Table III: \, \emph{Definitions of the parameters occurring in the \\ general 5-dimensional solutions of Lovelock and $AdS_5$ Chern--Simons gravity}
\end{center}
\begin{center}
\begin{tabular}{|c |c |c|}\hline\hline
\multicolumn{1}{|c|} \quad  &\textbf{ Lovelock Gravity} & \textbf{Chern--Simons Gravity} \\ \hline 
 $w$ & $  \displaystyle 3 \alpha_1^2 - 2 \alpha_0  \alpha_2$ &  $   \displaystyle \frac{14}{15 l^4}$ \\ 
  & & \\
 \cline{1-3}
 & & \\
 $z$ &$  \displaystyle \sqrt{c_1 \alpha_0 + 9\alpha_1^2}$ & $   \displaystyle \sqrt{\frac{20 + c_1}{5l^4}}$ \\ 
  & & \\
  \cline{1-3} 
  & & \\
 $u$ &$  \displaystyle \sqrt{c_1 \alpha_0 \alpha_1^2 + 6 \alpha_1^2 w -   2 \alpha_1 z \, w + 4 \alpha_0^2 \alpha_2^2}$ &  $   \displaystyle \frac{2}{15}\sqrt{\frac{149+5 c_1 - 14 \sqrt{5} \sqrt{20 + c_1}}{l^8}}$ \\ 
  & & \\
  \cline{1-3}
 & & \\
$ y$ & $  \displaystyle \frac{(-z \alpha_1 + w)}{u} $& $   \displaystyle \frac{7 - \sqrt{5} \sqrt{20+c_1}}{\sqrt{149+5 c_1 - 14 \sqrt{5} \sqrt{20 + c_1}}}$ \\ 
 & & \\
 \cline{1-3}
& & \\
 $s$ & $  \displaystyle -\frac{z \, \alpha_1 + w}{v}$ & $   \displaystyle \frac{-7 - \sqrt{5} \sqrt{20+c_1}}{\sqrt{149+5 c_1 + 14 \sqrt{5} \sqrt{20 + c_1}}} $\\ 
  & & \\
  \cline{1-3}
& & \\
 $v$ & $  \displaystyle \sqrt{c_1 \alpha_0 \alpha_1^2 + 6 \alpha_1^2 w +  2 \alpha_1 z\, w + 4 \alpha_0^2 \alpha_2^2}$ & $   \displaystyle \frac{2}{15}\sqrt{\frac{149+5 c_1 + 14 \sqrt{5} \sqrt{20 + c_1}}{l^8}}$ \\ 
  & & \\
  \cline{1-3}
 & & \\
 $x$ &$  \displaystyle 4 \alpha_2 (c_1 + 6 \alpha_2) $&$   \displaystyle 4 (c_1 + 6)$ \\
  & & \\
\hline
\end{tabular}
\end{center}
\end{appendix}


\begin{thebibliography}{99}

\bibitem{Joyce:2014kja}
A.~Joyce, B.~Jain, J.~Khoury and M.~Trodden,
``Beyond the Cosmological Standard Model,''
Phys. Rept. \textbf{568}, 1-98 (2015)
[arXiv:1407.0059 [astro-ph.CO]].

\bibitem{Koyama:2015vza}
K.~Koyama,
``Cosmological Tests of Modified Gravity,''
Rept. Prog. Phys. \textbf{79}, no.4, 046902 (2016)
[arXiv:1504.04623 [astro-ph.CO]].

\bibitem{Capozziello:2010zz}
 S.~Capozziello and V. Faraoni, 
``Beyond Einstein Gravity: A Survey of Gravitational Theories for Cosmology and Astrophysics,''
Fundam. Theor. Phys. \textbf{170} (2010)

\bibitem{Capozziello2002}
S.~Capozziello,
``Curvature quintessence,''
Int. J. Mod. Phys. D \textbf{11}, 483-492 (2002)
[arXiv:gr-qc/0201033 [gr-qc]].
  
\bibitem{Capozziello:2009nq}
S.~Capozziello, M.~De Laurentis and V.~Faraoni,
``A Bird's eye view of f(R)-gravity,''
Open Astron. J. \textbf{3}, 49 (2010)
[arXiv:0909.4672 [gr-qc]].

\bibitem{Nojiri:2017qvx}
S.~Nojiri, S.~D.~Odintsov and V.~K.~Oikonomou,
``Constant-roll Inflation in $F(R)$ Gravity,''
Class. Quant. Grav. \textbf{34}, no.24, 245012 (2017)
[arXiv:1704.05945 [gr-qc]].
  
\bibitem{Capozziello:2011et}
S.~Capozziello and M.~De Laurentis,
``Extended Theories of Gravity,''
Phys. Rept. \textbf{509}, 167-321 (2011)
[arXiv:1108.6266 [gr-qc]].

\bibitem{Capozziello:2015hra}
S.~Capozziello, G.~Gionti, S.J. and D.~Vernieri,
``String duality transformations in $f(R)$ gravity from Noether symmetry approach,''
JCAP \textbf{01}, 015 (2016)
[arXiv:1508.00441 [gr-qc]].

\bibitem{Ribeiro:2021gds} 
A.~R.~Ribeiro, D.~Vernieri and F.~S.~N.~Lobo, 
``Effective $f(R)$ actions for modified Loop Quantum Cosmologies via order reduction,'' [arXiv:2104.12283 [gr-qc]]. 

\bibitem{Capozziello:2014ioa}
S.~Capozziello, M.~De Laurentis and S.~D.~Odintsov,
``Noether Symmetry Approach in Gauss-Bonnet Cosmology,''
Mod. Phys. Lett. A \textbf{29}, no.30, 1450164 (2014)
[arXiv:1406.5652 [gr-qc]].

\bibitem{Terrucha:2019jpm}
I.~Terrucha, D.~Vernieri and J.~P.~S.~Lemos,
``Covariant action for bouncing cosmologies in modified Gauss\textendash{}Bonnet gravity,''
Annals Phys. \textbf{404}, 39-46 (2019)
[arXiv:1904.00260 [gr-qc]].

\bibitem{Barros:2019pvc}
B.~J.~Barros, E.~M.~Teixeira and D.~Vernieri,
``Bouncing cosmology in $f(R,\mathcal{G})$ gravity by order reduction,''
Annals Phys. \textbf{419}, 168231 (2020)
[arXiv:1907.11732 [gr-qc]].

\bibitem{Bajardi:2020mdp}
F.~Bajardi, S.~Capozziello and D.~Vernieri,
``Non-Local Curvature and Gauss-Bonnet Cosmologies by Noether Symmetries,''
Eur. Phys. J. Plus \textbf{135}, 942 (2020)
[arXiv:2011.01317 [gr-qc]].

\bibitem{Blazquez-Salcedo:2016enn}
J.~L.~Bl\'azquez-Salcedo, C.~F.~B.~Macedo, V.~Cardoso, V.~Ferrari, L.~Gualtieri, F.~S.~Khoo, J.~Kunz and P.~Pani,
``Perturbed black holes in Einstein-dilaton-Gauss-Bonnet gravity: Stability, ringdown, and gravitational-wave emission,''
Phys. Rev. D \textbf{94}, no.10, 104024 (2016)
[arXiv:1609.01286 [gr-qc]].

\bibitem{Blazquez-Salcedo:2017txk}
J.~L.~Bl\'azquez-Salcedo, F.~S.~Khoo and J.~Kunz,
``Quasinormal modes of Einstein-Gauss-Bonnet-dilaton black holes,''
Phys. Rev. D \textbf{96}, no.6, 064008 (2017)
[arXiv:1706.03262 [gr-qc]].

\bibitem{Amendola:1999qq}
L.~Amendola,
``Scaling solutions in general nonminimal coupling theories,''
Phys. Rev. D \textbf{60}, 043501 (1999)
[arXiv:astro-ph/9904120 [astro-ph]].

\bibitem{Uzan:1999ch}
J.~P.~Uzan,
``Cosmological scaling solutions of nonminimally coupled scalar fields,''
Phys. Rev. D \textbf{59}, 123510 (1999)
[arXiv:gr-qc/9903004 [gr-qc]].

\bibitem{Halliwell:1986ja}
J.~J.~Halliwell,
``Scalar Fields in Cosmology with an Exponential Potential,''
Phys. Lett. B \textbf{185}, 341 (1987)

\bibitem{Bajardi:2020xfj}
F.~Bajardi and S.~Capozziello,
``Equivalence of nonminimally coupled cosmologies by Noether symmetries,''
Int. J. Mod. Phys. D \textbf{29}, no.14, 2030015 (2020)
[arXiv:2010.07914 [gr-qc]].

\bibitem{Capozziello:2007ec}
S.~Capozziello and M.~Francaviglia,
``Extended Theories of Gravity and their Cosmological and Astrophysical Applications,''
Gen. Rel. Grav. \textbf{40}, 357-420 (2008)
[arXiv:0706.1146 [astro-ph]].

\bibitem{Clifton:2011jh}
T.~Clifton, P.~G.~Ferreira, A.~Padilla and C.~Skordis,
``Modified Gravity and Cosmology,''
Phys. Rept. \textbf{513}, 1-189 (2012)
[arXiv:1106.2476 [astro-ph.CO]].

\bibitem{Bamba:2012cp}
K.~Bamba, S.~Capozziello, S.~Nojiri and S.~D.~Odintsov,
``Dark energy cosmology: the equivalent description via different theoretical models and cosmography tests,''
Astrophys. Space Sci. \textbf{342}, 155-228 (2012)
[arXiv:1205.3421 [gr-qc]].

\bibitem{Nojiri:2017ncd}
S.~Nojiri, S.~D.~Odintsov and V.~K.~Oikonomou,
``Modified Gravity Theories on a Nutshell: Inflation, Bounce and Late-time Evolution,''
Phys. Rept. \textbf{692}, 1-104 (2017)
[arXiv:1705.11098 [gr-qc]].

\bibitem{Capozziello:2019klx}
S.~Capozziello and F.~Bajardi,
``Gravitational waves in modified gravity,''
Int. J. Mod. Phys. D \textbf{28}, no.05, 1942002 (2019)

\bibitem{Mantica}
S.~Capozziello, C.~A.~Mantica and L.~G.~Molinari,
``Cosmological perfect-fluids in f(R) gravity,''
Int. J. Geom. Meth. Mod. Phys. \textbf{16}, no.01, 1950008 (2018)
[arXiv:1810.03204 [gr-qc]].


\bibitem{Bajardi:2020fxh}
F.~Bajardi, D.~Vernieri and S.~Capozziello,
``Bouncing Cosmology in f(Q) Symmetric Teleparallel Gravity,''
Eur. Phys. J. Plus \textbf{135}, no.11, 912 (2020)
[arXiv:2011.01248 [gr-qc]].

\bibitem{Hartle:1983ai}
J.~B.~Hartle and S.~W.~Hawking,
``Wave Function of the Universe,''
Adv. Ser. Astrophys. Cosmol. \textbf{3}, 174-189 (1987)

\bibitem{Hawking:1983hj}
S.~W.~Hawking,
``The Quantum State of the Universe,''
Adv. Ser. Astrophys. Cosmol. \textbf{3}, 236-255 (1987)

\bibitem{Cai:2015emx}
Y.~F.~Cai, S.~Capozziello, M.~De Laurentis and E.~N.~Saridakis,
``f(T) teleparallel gravity and cosmology,''
Rept. Prog. Phys. \textbf{79}, no.10, 106901 (2016)
[arXiv:1511.07586 [gr-qc]].

\bibitem{Ferraro:2006jd}
R.~Ferraro and F.~Fiorini,
``Modified teleparallel gravity: Inflation without inflaton,''
Phys. Rev. D \textbf{75}, 084031 (2007)
[arXiv:gr-qc/0610067 [gr-qc]].
 
\bibitem{Arcos:2005ec}
H.~I.~Arcos and J.~G.~Pereira,
``Torsion gravity: A Reappraisal,''
Int. J. Mod. Phys. D \textbf{13}, 2193-2240 (2004)
[arXiv:gr-qc/0501017 [gr-qc]].

\bibitem{Bajardi:2021tul}
F.~Bajardi and S.~Capozziello,
``Noether symmetries and quantum cosmology in extended teleparallel gravity,''
Int. J. Geom. Meth. Mod. Phys. \textbf{18}, no.supp01, 2140002 (2021)

\bibitem{Lovelock:1971yv}
D.~Lovelock,
``The Einstein tensor and its generalizations,''
J. Math. Phys. \textbf{12}, 498-501 (1971)

\bibitem{Zanelli:2005sa}
J.~Zanelli,
``Lecture notes on Chern-Simons (super-)gravities. Second edition (February 2008),''
[arXiv:hep-th/0502193 [hep-th]].

\bibitem{Achucarro:1987vz}
A.~Achucarro and P.~K.~Townsend,
``A Chern-Simons Action for Three-Dimensional anti-De Sitter Supergravity Theories,''
Phys. Lett. B \textbf{180}, 89 (1986)

\bibitem{Chapline:1982ww}
G.~F.~Chapline and N.~S.~Manton,
``Unification of Yang-Mills Theory and Supergravity in Ten-Dimensions,''
Phys. Lett. B \textbf{120}, 105-109 (1983)

\bibitem{Benna:2008zy}
M.~Benna, I.~Klebanov, T.~Klose and M.~Smedback,
``Superconformal Chern-Simons Theories and AdS(4)/CFT(3) Correspondence,''
JHEP \textbf{09}, 072 (2008)
[arXiv:0806.1519 [hep-th]].

\bibitem{Chamseddine:1990gk}
A.~H.~Chamseddine,
``Topological gravity and supergravity in various dimensions,''
Nucl. Phys. B \textbf{346}, 213-234 (1990)

\bibitem{Aviles:2016hnm}
L.~Avil\'es, P.~Mella, C.~Quinzacara and P.~Salgado,
``Some cosmological solutions in Einstein-Chern-Simons gravity,'' [arXiv:1607.07137 [gr-qc]].

\bibitem{Gomez:2011zzd}
F.~Gomez, P.~Minning and P.~Salgado,
``Standard cosmology in Chern-Simons gravity,''
Phys. Rev. D \textbf{84}, 063506 (2011)

\bibitem{Regge:1986nz}
T.~Regge,
``On Broken Symmetries and Gravity,''
Phys. Rept. \textbf{137}, 31-33 (1986)

\bibitem{Cvetkovic:2016ios}
B.~Cvetkovi\'c and D.~Simi\'c,
``5D Lovelock gravity: new exact solutions with torsion,''
Phys. Rev. D \textbf{94}, no.8, 084037 (2016)
[arXiv:1608.07976 [gr-qc]].

\bibitem{Mardones:1990qc}
A.~Mardones and J.~Zanelli,
``Lovelock-Cartan theory of gravity,''
Class. Quant. Grav. \textbf{8}, 1545-1558 (1991)

\bibitem{Exirifard:2007da}
Q.~Exirifard and M.~M.~Sheikh-Jabbari,
``Lovelock gravity at the crossroads of Palatini and metric formulations,''
Phys. Lett. B \textbf{661}, 158-161 (2008)
[arXiv:0705.1879 [hep-th]].

\bibitem{Deruelle:2003ck}
N.~Deruelle and J.~Madore,
``On the quasilinearity of the Einstein-'Gauss-Bonnet' gravity field equations,''
[arXiv:gr-qc/0305004 [gr-qc]].

\bibitem{Boulware:1985wk}
D.~G.~Boulware and S.~Deser,
``String Generated Gravity Models,''
Phys. Rev. Lett. \textbf{55}, 2656 (1985)

\bibitem{Castillo-Felisola:2016kpe}
O.~Castillo-Felisola, C.~Corral, S.~del Pino and F.~Ram\'\i{}rez,
``Kaluza-Klein cosmology from five-dimensional Lovelock-Cartan theory,''
Phys. Rev. D \textbf{94}, no.12, 124020 (2016)
[arXiv:1609.09045 [gr-qc]].

\bibitem{Teitelboim:1987zz}
C.~Teitelboim and J.~Zanelli,
``Dimensionally continued topological gravitation theory in Hamiltonian form,''
Class. Quant. Grav. \textbf{4}, L125 (1987)

\bibitem{Miskovic:2007mg}
O.~Miskovic and R.~Olea,
``Counterterms in Dimensionally Continued AdS Gravity,''
JHEP \textbf{10}, 028 (2007)
[arXiv:0706.4460 [hep-th]].

\bibitem{Miskovic:2010ui}
O.~Miskovic and R.~Olea,
``Conserved charges for black holes in Einstein-Gauss-Bonnet gravity coupled to nonlinear electrodynamics in AdS space,''
Phys. Rev. D \textbf{83}, 024011 (2011)
[arXiv:1009.5763 [hep-th]].

\bibitem{Kofinas:2008ub}
G.~Kofinas and R.~Olea,
``Universal Kounterterms in Lovelock AdS gravity,''
Fortsch. Phys. \textbf{56}, 957-963 (2008)
[arXiv:0806.1197 [hep-th]].

\bibitem{Oliva:2010zd}
J.~Oliva and S.~Ray,
``Classification of Six Derivative Lagrangians of Gravity and Static Spherically Symmetric Solutions,''
Phys. Rev. D \textbf{82}, 124030 (2010)
[arXiv:1004.0737 [gr-qc]].

\bibitem{Green:1987sp}
  M.~B.~Green, J.~H.~Schwarz and E.~Witten,
 `` Superstring Theory. Vol. 1: Introduction," Cambridge Monographs On Mathematical Physics, Cambridge, UK: Univ. Pr. (1987).
  
\bibitem{Green:1987mn}
  M.~B.~Green, J.~H.~Schwarz and E.~Witten,
 `` Superstring Theory. Vol. 2: Loop Amplitudes, Anomalies And Phenomenology," Cambridge Monographs On Mathematical Physics, Cambridge, UK: Univ. Pr. (1987).
  
\bibitem{Polchinski:1998rq}
  J.~Polchinski,
  ``String theory. Vol. 1: An introduction to the bosonic string," Cambridge, UK: Univ. Pr. (1998).
  
\bibitem{Polchinski:1998rr}
  J.~Polchinski,
 `` String theory. Vol. 2: Superstring theory and beyond," Cambridge, UK: Univ. Pr. (1998).
  
\bibitem{Becker:2007zj}
  K.~Becker, M.~Becker and J.~H.~Schwarz,
  ``String theory and M-theory: A modern introduction," Cambridge, UK: Univ. Pr. (2006).
  
\bibitem{Dine:2000bf}
M.~Dine,
``TASI lectures on M theory phenomenology,''
doi:10.1142/9789812799630\_0006

\bibitem{Witten:1995em}
E.~Witten,
``Five-branes and M theory on an orbifold,''
Nucl. Phys. B \textbf{463}, 383-397 (1996)
  
\bibitem{Klein:1926tv}
O.~Klein,
``Quantum Theory and Five-Dimensional Theory of Relativity. (In German and English),''
Z. Phys. \textbf{37}, 895-906 (1926)
  
\bibitem{Kaluza:1921tu}
T.~Kaluza,
``Zum Unit\"atsproblem der Physik,''
Sitzungsber. Preuss. Akad. Wiss. Berlin (Math. Phys. ) \textbf{1921}, 966-972 (1921)
[arXiv:1803.08616 [physics.hist-ph]].
  
\bibitem{Han:1998sg}
T.~Han, J.~D.~Lykken and R.~J.~Zhang,
``On Kaluza-Klein states from large extra dimensions,''
Phys. Rev. D \textbf{59}, 105006 (1999)
[arXiv:hep-ph/9811350 [hep-ph]].

\bibitem{Mueller-Hoissen:1985www}
F.~Mueller-Hoissen,
``Dimensionally Continued Euler Forms, {Kaluza-Klein} Cosmology and Dimensional Reduction,''
Class. Quant. Grav. \textbf{3}, 665 (1986)

\bibitem{Crisostomo:2000bb}
J.~Crisostomo, R.~Troncoso and J.~Zanelli,
``Black hole scan,''
Phys. Rev. D \textbf{62}, 084013 (2000)
[arXiv:hep-th/0003271 [hep-th]].

\bibitem{Zwiebach:1985uq}
B.~Zwiebach,
``Curvature Squared Terms and String Theories,''
Phys. Lett. B \textbf{156}, 315-317 (1985)

\bibitem{Zumino:1985dp}
B.~Zumino,
``Gravity Theories in More Than Four-Dimensions,''
Phys. Rept. \textbf{137}, 109 (1986)

\bibitem{DeFelice:2009ak}
A.~De Felice and T.~Suyama,
``Vacuum structure for scalar cosmological perturbations in Modified Gravity Models,''
JCAP \textbf{06}, 034 (2009)
[arXiv:0904.2092 [astro-ph.CO]].

\bibitem{Astashenok:2015haa}
A.~V.~Astashenok, S.~D.~Odintsov and V.~K.~Oikonomou,
``Modified Gauss\textendash{}Bonnet gravity with the Lagrange multiplier constraint as mimetic theory,''
Class. Quant. Grav. \textbf{32}, no.18, 185007 (2015)
[arXiv:1504.04861 [gr-qc]].

\bibitem{Nojiri:2018ouv}
S.~Nojiri, S.~D.~Odintsov and V.~K.~Oikonomou,
``Ghost-free Gauss-Bonnet Theories of Gravity,''
Phys. Rev. D \textbf{99}, no.4, 044050 (2019)
[arXiv:1811.07790 [gr-qc]].

\bibitem{Wheeler:1985nh}
J.~T.~Wheeler,
``Symmetric Solutions to the Gauss-Bonnet Extended Einstein Equations,''
Nucl. Phys. B \textbf{268}, 737-746 (1986)

\bibitem{Bajardi:2020osh}
F.~Bajardi and S.~Capozziello,
``$f(\mathcal {G})$ Noether cosmology,''
Eur. Phys. J. C \textbf{80}, no.8, 704 (2020)
[arXiv:2005.08313 [gr-qc]].

\bibitem{Deruelle:1989fj}
N.~Deruelle and L.~Farina-Busto,
``The Lovelock Gravitational Field Equations in Cosmology,''
Phys. Rev. D \textbf{41}, 3696 (1990)

\bibitem{Deruelle:1987ux}
N.~Deruelle and J.~Madore,
``A smooth oscillating cosmological solution,''
Phys. Lett. B \textbf{186}, 25-28 (1987)

\bibitem{Myers:1988ze}
R.~C.~Myers and J.~Z.~Simon,
``Black Hole Thermodynamics in Lovelock Gravity,''
Phys. Rev. D \textbf{38}, 2434-2444 (1988)

\bibitem{Aros:2000ij}
R.~Aros, R.~Troncoso and J.~Zanelli,
``Black holes with topologically nontrivial AdS asymptotics,''
Phys. Rev. D \textbf{63}, 084015 (2001)
[arXiv:hep-th/0011097 [hep-th]].

\bibitem{Myers:1989kt}
R.~C.~Myers and J.~Z.~Simon,
``Black Hole Evaporation and Higher Derivative Gravity,''
Gen. Rel. Grav. \textbf{21}, 761-766 (1989)

\bibitem{Jacobson:1993vj}
T.~Jacobson, G.~Kang and R.~C.~Myers,
``On black hole entropy,''
Phys. Rev. D \textbf{49}, 6587-6598 (1994)
[arXiv:gr-qc/9312023 [gr-qc]].

\bibitem{Jacobson:1993xs}
T.~Jacobson and R.~C.~Myers,
``Black hole entropy and higher curvature interactions,''
Phys. Rev. Lett. \textbf{70}, 3684-3687 (1993)
[arXiv:hep-th/9305016 [hep-th]].

\bibitem{Ahmed:2014eea}
N.~Ahmed and H.~Rafat,
``Deformation Retract and Folding of the 5D Schwarzchild Field,''
[arXiv:1405.1057 [physics.gen-ph]].

\bibitem{Myung:2001ab}
Y.~S.~Myung,
``Entropy of the three-dimensional Schwarzschild-de Sitter black hole,''
Mod. Phys. Lett. A \textbf{16}, 2353 (2001)
[arXiv:hep-th/0110123 [hep-th]].

\bibitem{Aharony:1998tt}
O.~Aharony and T.~Banks,
``Note on the quantum mechanics of M theory,''
JHEP \textbf{03}, 016 (1999)
[arXiv:hep-th/9812237 [hep-th]].

\bibitem{Garraffo:2008hu}
C.~Garraffo and G.~Giribet,
``The Lovelock Black Holes,''
Mod. Phys. Lett. A \textbf{23}, 1801-1818 (2008)
[arXiv:0805.3575 [gr-qc]].

\bibitem{Anabalon:2009kq}
A.~Anabalon, N.~Deruelle, Y.~Morisawa, J.~Oliva, M.~Sasaki, D.~Tempo and R.~Troncoso,
``Kerr-Schild ansatz in Einstein-Gauss-Bonnet gravity: An exact vacuum solution in five dimensions,''
Class. Quant. Grav. \textbf{26}, 065002 (2009)
[arXiv:0812.3194 [hep-th]].

\bibitem{Banados:1993ur}
M.~Banados, C.~Teitelboim and J.~Zanelli,
``Dimensionally continued black holes,''
Phys. Rev. D \textbf{49}, 975-986 (1994)
[arXiv:gr-qc/9307033 [gr-qc]].

\bibitem{Wiltshire:1988uq}
D.~L.~Wiltshire,
``Black Holes in String Generated Gravity Models,''
Phys. Rev. D \textbf{38}, 2445 (1988)

\bibitem{Whitt:1988ax}
B.~Whitt,
``Spherically Symmetric Solutions of General Second Order Gravity,''
Phys. Rev. D \textbf{38}, 3000 (1988)

\bibitem{Bajardi:2019zzs}
F.~Bajardi, K.~F.~Dialektopoulos and S.~Capozziello,
``Higher Dimensional Static and Spherically Symmetric Solutions in Extended Gauss\textendash{}Bonnet Gravity,''
Symmetry \textbf{12}, no.3, 372 (2020)
[arXiv:1911.03554 [gr-qc]].

\bibitem{Mukherjee:2020lld}
S.~Mukherjee and N.~Dadhich,
``Pure Gauss\textendash{}Bonnet NUT black hole with and without non-central singularity,''
Eur. Phys. J. C \textbf{81}, no.5, 458 (2021)
[arXiv:2012.15560 [gr-qc]].

\bibitem{Mukherjee:2021erg}
S.~Mukherjee and N.~Dadhich,
``Pure Gauss-Bonnet NUT Black Hole Solution: I,''
[arXiv:2101.02958 [gr-qc]].

\bibitem{Glavan:2019inb}
D.~Glavan and C.~Lin,
``Einstein-Gauss-Bonnet Gravity in Four-Dimensional Space-time,''
Phys. Rev. Lett. \textbf{124}, no.8, 081301 (2020)
[arXiv:1905.03601 [gr-qc]].

\bibitem{Gurses:2020rxb}
M.~Gurses, T.~\c{C}.~\c{S}i\c{s}man and B.~Tekin,
``Comment on ''Einstein-Gauss-Bonnet Gravity in 4-Dimensional Space-Time'',''
Phys. Rev. Lett. \textbf{125}, no.14, 149001 (2020)
[arXiv:2009.13508 [gr-qc]].
\end{thebibliography}
\end{document}